\documentclass[pra,aps,superscriptaddress,nofootinbib,twocolumn]{revtex4-1}

\usepackage{mathrsfs}
\usepackage{amsfonts}
\usepackage{amssymb}
\usepackage{amsmath}
\usepackage{graphicx}
\usepackage[usenames,dvipsnames]{color}
\usepackage[colorlinks=true,citecolor=
blue,linkcolor=magenta]{hyperref}
\usepackage{xcolor}
\usepackage{ulem}
\usepackage{lmodern}
\usepackage{amsthm}
\usepackage[utf8]{inputenc}

\newcommand{\figpath}{.}

\newcommand{\Tr}{\mathrm{Tr}}

\newcommand{\abs}[1]{\vert #1 \vert}
\newcommand{\absLR}[1]{\left\vert #1 \right\vert}

\newcommand{\ket}[1]{\vert{ #1 }\rangle}
\newcommand{\bra}[1]{\langle{ #1 }\vert}
\newcommand{\ketbra}[2]{\vert #1 \rangle \langle #2 \vert}

\newcommand{\mean}[1]{\langle #1 \rangle}

\newcommand{\st}{\,\vert\,}

\newcommand{\bfR}{\boldsymbol{R}}
\newcommand{\bfP}{\boldsymbol{P}}
\newcommand{\bfq}{\boldsymbol{q}}
\newcommand{\bfI}{\boldsymbol{I}}
\newcommand{\bfmu}{\boldsymbol{\mu}}
\newcommand{\bfsigma}{\boldsymbol{\sigma}}
\newcommand{\bfa}{\boldsymbol{a}}
\newcommand{\bfb}{\boldsymbol{b}}
\newcommand{\bfzero}{\boldsymbol{0}}
\newcommand{\bfp}{\boldsymbol{p}}

\newcommand{\calG}{\mathcal{G}}
\newcommand{\calR}{\mathcal{R}}
\newcommand{\calP}{\mathcal{P}}
\newcommand{\calM}{\mathcal{M}}
\newcommand{\calS}{\mathcal{S}}
\newcommand{\calF}{\mathcal{F}}
\newcommand{\calN}{\mathcal{N}}
\newcommand{\calD}{\mathcal{D}}

\begin{document}

\title{Learning-based quantum error mitigation}

\author{Armands Strikis}
\thanks{These two authors contributed equally.}
\affiliation{Department of Materials, University of Oxford, Oxford OX1 3PH, United Kingdom}

\author{Dayue Qin}
\thanks{These two authors contributed equally.}
\affiliation{Graduate School of China Academy of Engineering Physics, Beijing 100193, China}

\author{Yanzhu Chen}
\affiliation{C. N. Yang Institute for Theoretical Physics, State University of New York at Stony Brook, Stony Brook, NY 11794-3840, USA}
\affiliation{Department of Physics and Astronomy, State University of New York at Stony Brook, Stony Brook, NY 11794-3800, USA}

\author{Simon C. Benjamin}
\affiliation{Department of Materials, University of Oxford, Oxford OX1 3PH, United Kingdom}

\author{Ying Li}
\email{yli@gscaep.ac.cn}
\affiliation{Graduate School of China Academy of Engineering Physics, Beijing 100193, China}

\begin{abstract}
If NISQ-era quantum computers are to perform useful tasks, they will need to employ powerful error mitigation techniques. Quasi-probability methods can permit perfect error compensation at the cost of additional circuit executions, provided that the nature of the error model is fully understood and sufficiently local both spatially and temporally. Unfortunately these conditions are challenging to satisfy. Here we present a method by which the proper compensation strategy can instead be learned {\it ab initio}. Our training process uses multiple variants of the primary circuit where all non-Clifford gates are substituted with gates that are efficient to simulate classically. The process yields a configuration that is near-optimal versus noise in the real system with its non-Clifford gate set. Having presented a range of learning strategies, we demonstrate the power of the technique both with real quantum hardware (IBM devices) and exactly-emulated imperfect quantum computers. The systems suffer a range of noise severities and types, including spatially and temporally correlated variants. In all cases the protocol successful adapts to the noise and mitigates it to a high degree.
\end{abstract}

\maketitle

\section{Introduction}

It is widely believed that we are entering the era when the computational power of quantum machines surpasses any classical resource for certain specific problems~\cite{Nielsen2010, Arute2019}. One of the main obstacles to achieving the practical application of quantum computing is the noise caused by decoherence and imperfect control. There exist well-understood solutions involving quantum error correction, which can suppress the computing error to an arbitrarily low-level when the error rate of elementary gates is lower than the threshold. However, implementing this approach involves a multiplicative increase in the number of physical qubits, potentially by a factor of a thousand or more~\cite{Fowler2012}. This appears prohibitive for the near future. Therefore for noisy intermediate-scale quantum (NISQ) devices, alternative approaches which are usually termed quantum error {\it mitigation} have been developed. 
 
At the base level, it is of course essential to minimise noise during the physical execution of a gate, through optimising control parameters et cetera~\cite{Motzoi2009}, and here we take it as read that such measures have been taken. Above this level, one can use error extrapolation and probabilistic error cancellation~\cite{Li2017, Temme2017, Endo2018}; here the estimator of the computing result is carefully constructed and optimised using the knowledge of error distribution, such that the impact of errors is minimised~\cite{Kandala2019, Song2019, Zhang2020}. Similar ideas have been used to correct measurement errors~\cite{Kwon2020, Chen2019, Geller2020}. By exploring the symmetry of the quantum circuit, some errors in the circuit can be detected and eliminated using post-selection~\cite{McArdle2019, Bonet2018}. A number of related ideas such as subspace expansion~\cite{McClean2017} and continuous error mitigation~\cite{Sun2020}, among others, are being explored. 

In many potential NISQ applications, for example the use of variational quantum algorithms (QVAs) in eigensolver or simulation~\cite{Peruzzo2014, Li2017, O'Malley2016}, a key task is to evaluate mean values of some observables -- in essence, to measure the expected value of one or more qubits as the output of a circuit. The estimator of the mean is usually biased as a result of the noise. Then the role of error mitigation is to remove the bias by modifying the estimator. Because of the linearity of quantum mechanics, a linear combination of noisy circuits with appropriate coefficients (both positive and negative) can be equivalent to a noise-free circuit~\cite{Temme2017}. One can implement such a combination by randomly sampling from a particular set of quantum circuits, derived from the primary circuit by (typically) the addition of certain gate(s), and taking a weighted average over the recorded outcomes. This can be called probabilistic error cancellation~\cite{Temme2017}. If the error model, i.e. a precise theoretical characterisation of the errors in the physical gates, is available then it may be possible to analytically derive the ideal distribution of circuits, both their nature and the proper weightings with which their outputs should be combined. Then perfect compensation for errors is achievable~\cite{Temme2017,Endo2018}. However, for this to be practical, the error model must be determined through some form of tomography~\cite{Endo2018}; this may be difficult~\cite{Wise2020} or infeasibly costly unless error correlations (either spatial or temporal) involve only a few qubits. Nevertheless, when such conditions are even approximately met then the approach can be very valuable, as has been successfully demonstrated in small systems of superconducting qubits and trapped ions~\cite{Song2019, Zhang2020}. 

In this paper, we present a novel and intuitive way to mitigate the errors. Instead of determining the error model that afflicts the experimental system and deriving the proper circuit distribution (i.e.~combination coefficients), the distribution is determined via an {\it ab initio} learning process. We choose the distribution by minimising the error in the final computing result for a set of training computing tasks. The efficiency of the learning-based error mitigation is due to its simplicity and intuitivity. All potential error correlations, i.e.~spatial and temporal correlations, are automatically taken into account in the learning process. Therefore, it is a promising way to realise reliable quantum computing with deep circuits on large systems.

An obvious difficulty for a learning-based error mitigation process, if it is to be relevant to real quantum computers implemented at scale, is that one cannot determine the correct value of a given observable (the `goal' of the mitigation) by any means other than the execution of an ideal quantum circuit! Here we show that learning-based error-mitigation is indeed feasible because Clifford-circuit {\it training tasks} are sufficient to find an optimal circuit distribution, regardless of the error correlations. We derive suitable Clifford circuits from the original (primary) circuit, and for such circuits we can evaluate the correct result by using efficient simulations on a classical computer~\cite{Gottesman1998, Aaronson2004, Anders2006}. We note that in the present work, the sufficiency is proved under the assumption of negligible single-qubit gate errors. In most quantum computing systems, single-qubit gates do indeed attain a much higher fidelity than other gates, e.g.~an average gate fidelity of $99.9999\%$ has been achieved with trapped ions~\cite{Harty2014} whereas the record for two-qubit fidelity is three orders of magnitude lower at $99.9\%$~\cite{Ballance2016, Gaebler2016}. 

In the following we will consider two types of quantum computers, and argue that they are practically equivalent. The distinction concerns the question of whether it is trivial (zero resource cost) to reconfigure the computer from one circuit to another. The more convenient theoretical assumption is that it is indeed cost-free to reconfigure, in which case the learning process is a structureless random sampling. In real systems an experimentalist may prefer to configure a circuit once and sample from it many times before reconfiguring. The learning-based error-mitigation has two stages: the learning, in which we need to evaluate a loss function, and the error-mitigated computation. If the quantum circuit can be updated after each run, we use the Monte Carlo summation in both the loss function evaluation and computing, in order to maximise the number of training circuits. For scenarios where reconfiguration is costly, we propose a method using significant-error interventions.

We demonstrate our protocol both with real quantum hardware and with exactly-simulated virtual devices. We consider various tasks that these devices are attempting to perform, including variational quantum algorithms. For the simulated machines we are of course able to specify the noise model as we wish.
 In order to compare with  previously reported tomography-based methods~\cite{Temme2017,Endo2018}, we specify that there are local (two-qubit) errors that conform to a known noise model but that the real noise model also involves additional correlated errors: either spatial `cross talk' or temporal correlations. The learning-based protocol outperforms the tomography-based protocol by a factor of approximately $4$-to-$5$ depending on the task at hand. Indeed in all cases that we explore, with the real or virtual quantum systems, we find that the learning-based protocol performs very well.

We comment and show numerical data for the scalability of our method to larger systems, and we consider various learning strategies (including single-parameter versus multi-parameter, and summation versus product ansatz) and we consider the distinction between ideal `infinite time' learning and resource-constrained learning. Because of the simplicity, effectiveness and flexibility of the learning-based approach, we conclude that it is a promising way to realise reliable value from NISQ-era quantum computing.

This paper is organised as follows. In Section~\ref{sec:generalP} and Section~\ref{sec:Properties} we introduce and describe the general protocol. In Section~\ref{sec:practical} we discuss practical implementation of the protocol, and focusing on a Pauli error model (Section~\ref{sec:Pauli}) we describe three practical methods in detail in Section~\ref{sec:SigE}, Section~\ref{sec:product-form} and Section~\ref{sec:Boltzmann} with numerical results in the first two. Section~\ref{sec:fidelity} separately introduces an alternative way to establish the cost function. In Section~\ref{sec:experiment} we demonstrate our protocol on real quantum hardware. Finally, in Section~\ref{sec:Conclusions} we summarise the protocol, conclude the main results and discuss future directions.

\section{The general protocol}
\label{sec:generalP}

\begin{figure*}[tbp]
\centering
\includegraphics[width=1\linewidth]{\figpath 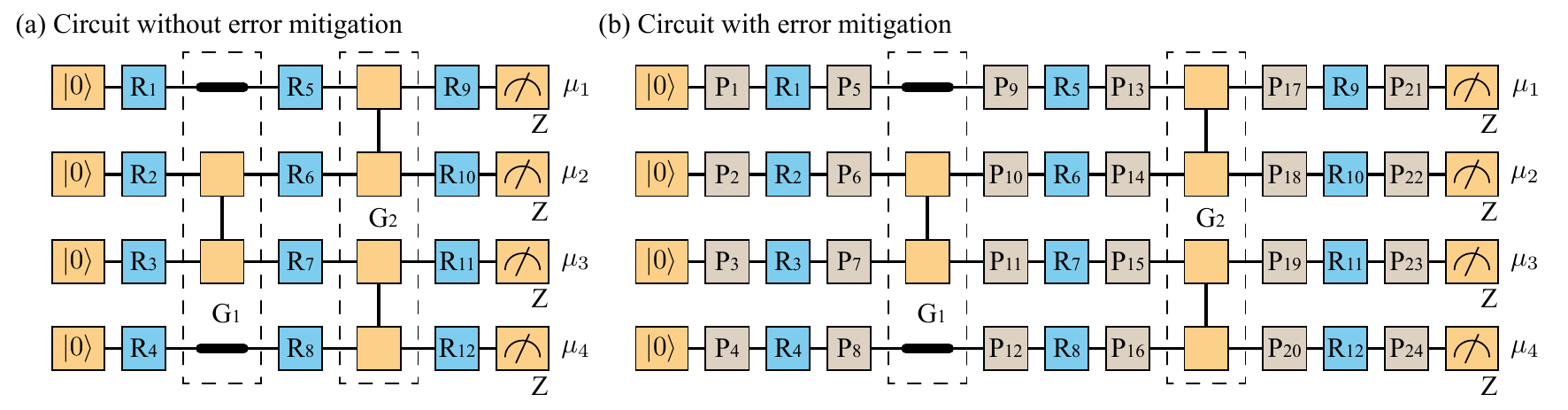}
\caption{
Simple example of circuit without and with error mitigation. Each circuit has four qubits and two layers of frame gates $G_1$ and $G_2$ (dashed boxes). Note that errors afflict the frame gates and may correlate over arbitrarily many qubits within a box and between boxes (i.e. spatial and temporal errors). Frame operations (orange) include the qubit initialisation, frame gates and measurement. There are three layers of computing gates $R_i$ (blue) in each circuit. To implement the error mitigation, two layers of Pauli gates $P_i$ (brown) are introduced before and after each layer of computing gates.  Usually, single-qubit gates next to each other in the circuit can be combined into one single-qubit gate in the physical implementation.
}
\label{fig:circuits}
\end{figure*}

We consider the quantum circuit as shown in Fig.~\ref{fig:circuits}. In the circuit, all qubits are initialised in the state $\ket{0}$ and measured in the $Z$ basis at the end. Most errors are caused by multi-qubit quantum gates, e.g.~controlled-NOT and controlled-phase gates. We call these gates frame gates. Suppose that the circuit has $n$ qubits and $N$ layers of frame gates, we use $G_j$, where $j=1,\ldots,N$ to denote the overall $n$-qubit gate for the $j$-th layer. We assume that these multi-qubit gates are all Clifford, which is the only requirement for frame gates. Between frame operations, single-qubit unitary gates $\bfR = (R_1,R_2,\ldots,R_{n(N+1)})$ are performed [see Fig.~\ref{fig:circuits}(a)], which specify the quantum computation. We call them computing gates. To implement the error mitigation, we introduce single-qubit Pauli gates before and after each computing gate [see Fig.~\ref{fig:circuits}(b)], which are denoted by $\bfP = (P_1,P_2,\ldots,P_{2n(N+1)})$. We call these Pauli gates error-mitigating gates. In our protocol, the frame gates $G_j$ are fixed, and other gates (i.e.~$\bfR$ and $\bfP$) are treated as variables. We remark that circuits composed in this way is universal for quantum computing, and our protocol can be generalised to other circuit configurations. 

Let $\bfmu$ be a binary vector that represents measurement outcomes of $n$ qubits. A specific computation is to evaluate the mean value of a function $f(\bfmu)$. For example, if the observable is $Z$ of the first qubit, the function is $f(\bfmu) = 1 - 2\mu_1$, where $\mu_1$ is the measurement outcome of the first qubit. We use ${\rm com}^{\rm ef}(\bfR,\bfP)$ to denote the mean value when the circuit is error-free and ${\rm com}(\bfR,\bfP)$ to denote the mean value in the actual noisy circuit. 

In probabilistic error mitigation, we use a linear combination of computing results with different $\bfP$ to estimate the error-free result. Given the combination coefficients $q(\bfP)$, i.e.~quasi-probabilities, the error-mitigated computing result is 
\begin{eqnarray}
\label{eq:errorMit}
{\rm com}^{\rm em}(\bfR,\bfI) \equiv \sum_{\bfP} q(\bfP) {\rm com}(\bfR,\bfP),
\end{eqnarray}
where $\bfI$ means that all error-mitigating gates are identity. Compared to the error-free result, the computing error is 
\begin{eqnarray}
{\rm Error}(\bfR) \equiv \absLR{ {\rm com}^{\rm em}(\bfR,\bfI) - {\rm com}^{\rm ef}(\bfR,\bfI) }.
\end{eqnarray}
Our goal is to find an optimal distribution $q(\bfP)$ such that the error is minimised. 

We consider the loss function in the quadratic form: 
\begin{eqnarray}
{\rm Loss} \equiv \frac{1}{\abs{\mathbb{T}}} \sum_{\bfR \in \mathbb{T}} {\rm Error}(\bfR)^2,
\end{eqnarray}
where $\mathbb{T}$ is a set of training computing tasks. To evaluate the loss function, we can compute ${\rm com}(\bfR,\bfI)$ using the actual noisy quantum computer and ${\rm com}^{\rm ef}(\bfR,\bfI)$ using a classical computer. Because Clifford circuits can be efficiently simulated on a classical computer according to the Gottesman-Knill theorem~\cite{Gottesman1998, Aaronson2004, Anders2006}, we choose the training set $\mathbb{T}$ as a subset of Clifford circuits, i.e.~$\mathbb{T} \subseteq \mathbb{C} \equiv \{ \bfR \st \text{All }R_j\text{ are Clifford} \}$. By minimising the loss function, we can find the optimal distribution $q_{\rm opt}(\bfP)$ for the training set. Then, we apply the same distribution $q_{\rm opt}(\bfP)$ to our primary computing task(s) $\bfR$. We remark that $\bfR$ will be non-Clifford in non-trivial quantum computations. 

\section{Key properties}
\label{sec:Properties}

An optimal distribution $q(\bfP)$ that works for all $\bfR$ exists if single-qubit gates are ideal. The error-free computation result can be expressed as (see Appendix~\ref{app:formalism}) 
\begin{eqnarray}
{\rm com}^{\rm ef}(\bfR,\bfP) = \Tr\left(\calS_{\rm S}\calR\calP_{\rm L}\calF^{\rm ef}\calP_{\rm R}\right).
\end{eqnarray}
Each term in trace brackets is a map on $n(N+1)$ qubits: $\calF^{\rm ef} = [G_N]\otimes\cdots\otimes[G_1]\otimes\calG_0^{\rm ef}$ is a tensor that describes the effect of all error-free frame operations, $\calR$ represents computing gates, $\calP_{\rm L}$ and $\calP_{\rm R}$ respectively represent error-mitigating gates in odd- and even-layers, and $\calS_{\rm S}$ is a swap map. Here, $[U](\bullet)=U\bullet U^\dag$ is the completely positive map of the operator $U$, $\calG_0^{\rm ef}(\bullet) = \rho_i^{\rm ef}\Tr\left(E_f^{\rm ef}\bullet\right)$ describes the qubit initialisation and measurement, and $\rho_i^{\rm ef}$ and $E_f^{\rm ef}$ are respectively the error-free initial state and measurement operator. The actual computation result with error can be expressed in the same form: 
\begin{eqnarray}
{\rm com}(\bfR,\bfP) = \Tr\left(\calS_{\rm S}\calR\calP_{\rm L}\calF\calP_{\rm R}\right),
\label{eq:com}
\end{eqnarray}
where $\calF$ describes the effect of all frame operations with errors. In general, $\calF$ cannot be written as a tensor product similar to $\calF^{\rm ef}$, specifically in the presence of correlated errors. The error-mitigated computation result is ${\rm com}^{\rm em}(\bfR,\bfI) = \Tr\left(\calS_{\rm S}\calR\calF^{\rm em}\right)$, where $\calF^{\rm em} = \sum_{\bfP} q(\bfP) \calP_{\rm L}\calF\calP_{\rm R}$. Therefore, the error is zero for all $\bfR$ if $q(\bfP)$ is a solution of the equation $\calF^{\rm em} = \calF^{\rm ef}$. The solution always exists if for every non-zero element of $\calF^{\rm ef}$ the corresponding element of $\calF$ is also non-zero in the the Pauli transfer matrix representation~\cite{Merkel2013, Greenbaum2015}. It is very unlikely that this condition does not hold, especially when the error rate is low. See Appendix~\ref{app:existence} for the proof. 

The training set $\mathbb{T} = \mathbb{C}$ is sufficient for finding an optimal distribution $q(\bfP)$ that works for all $\bfR$. The set $\mathbb{C}$ contains all Clifford $\bfR$. A single-qubit unitary map $[R_j]$ can be written as a linear combination of single-qubit Clifford maps~\cite{Endo2018}. For an arbitrary $\bfR$, we have $\calR = \sum_{\bfR'\in\mathbb{C}} \alpha_{\bfR,\bfR'}\calR'$, where $\alpha_{\bfR,\bfR'}$ are coefficients. See Appendix~\ref{app:completeness} for details. Therefore, if we find the optimal distribution $q_{\rm opt}(\bfP)$ such that ${\rm Loss} = 0$ with $\mathbb{T} = \mathbb{C}$, the error is zero for all Clifford and non-Clifford $\bfR$ after the error mitigation. 

We remark that these two properties are proved under the condition of ideal single-qubit unitary gates but do not depend on the error model of frame operations. When single-qubit-gate errors are gate-independent, the proofs still hold after some adaptation. The protocol works for all Pauli, damping and coherent, uncorrelated and correlated errors. 

\section{Practical issues}
\label{sec:practical}

The spaces of computing gates $\bfR$ and error-mitigating gates $\bfP$ increase exponentially with the circuit size. Therefore, it is impractical to compute ${\rm Error}(\bfR)$ for every training circuit $\bfR\in \mathbb{C}$ and optimise the quasi-probability $q(\bfP)$ of each $\bfP$. There are two approaches for the practical implementation as follows. 

In the first approach, we truncate spaces of training circuits and error-mitigating gates. We then require some rationale for choosing truncated sets that can be expected to be effective. This can be called the significant-error approach. An effective approach is to consider the Pauli error model. General errors can be converted into Pauli errors using the Pauli twirling, which will be discussed later. Pauli errors are erroneous Pauli gates. Usually only a small subset of Pauli errors are significant, which can be corrected by corresponding error-mitigating gates. Let ${\rm SigE}$ be the set of significant Pauli errors, we can take quasi-probabilities $q(\bfP)\vert_{\bfP\in{\rm SigE}}$ as optimisation parameters and set the rest $q(\bfP)\vert_{\bfP\notin{\rm SigE}}$ to zero. Then, the number of optimisation parameters is the same as the number of significant errors, which usually increases polynomially with the circuit size. Similarly, we choose a selected subset $\mathbb{T} \subset \mathbb{C}$ as the training set. Later, we will show the numerical evidence that the error mitigation works well when the size of $\mathbb{T}$ is three times the size of ${\rm SigE}$.


In the second approach, instead of truncating the space of training circuits and error-mitigating gates, we consider an error ansatz whose distribution admits a product form. That is, we consider a case where each of significant Pauli errors has its own independent quasi-probability distribution that we optimise. Then an application of error mitigation consists of applying chains of individual significant errors and has a corresponding quasi-probability distribution described as a product of independent quasi-probabilities for each significant error.

Finally, we may generalise the previous approach. We parameterise the quasi-probability distribution as a variational function and compute the loss using the Monte Carlo method. We take $q(\bfP) \propto B(\bfP,\lambda)$, where $\lambda$ denotes a set of parameters that determine the distribution. Here, $B(\bfP,\lambda)$ can be any real-valued function describing the distribution on the large space of $\bfP$ but only using a relatively small number of parameters $\lambda$, e.g.~the restricted Boltzmann machine~\cite{Fischer2012, Carleo2017}. Instead of the truncated training set, we can use the full set of Clifford circuits, i.e.~$\mathbb{T} = \mathbb{C}$. The loss function can be efficiently computed using the Monte Carlo summation. We find that the sampling cost scales polynomially with respect to the accuracy of the Monte Carlo summation regardless of the size of $\mathbb{C}$ and the space of $\bfP$. 

All three approaches will be discussed in this paper. We note that one can combine the approaches in different ways in a practical implementation. For example, we can use the significant-error approach to parameterise the distribution and use the Monte Carlo summation to evaluate the loss function. Having obtained an optimised quasi-probability distribution $q_{\rm opt}(\bfP)$ by any such method, we can implement the error-mitigated computation by using either the truncated space of error-mitigating gates or the Monte Carlo method. 

\section{Multiple observables and fidelity loss}
\label{sec:fidelity}

So far, we only considered the case of one observable $f(\bfmu)$. In some algorithms, e.g.~the variational quantum eigensolver~\cite{Peruzzo2014}, we need to measure multiple observables. The loss function can be generalised accordingly. Let ${\rm Loss}_f$ be the loss of the observable $f(\bfmu)$. Then, we can take the loss of $N_{\rm o}$ observables as $\overline{\rm Loss} \equiv N_{\rm o}^{-1}\sum_{i=1}^{N_{\rm o}} {\rm Loss}_{f_i}$, where $f_i(\bfmu)$ is the $i$-th observable. 

A further option is to base the cost function on the output state fidelity, a measure of the correctness that is independent of the observable. We can also use the fidelity to find an optimal quasi-probability distribution. Let $\ket{\psi(\bfR)}$ be the ideal final state (just before the measurement) of the circuit with gate sequences $\bfR$ and $\bfP = \bfI$. The quadratic fidelity loss function reads 
\begin{eqnarray}
\widetilde{\rm Loss} \equiv \frac{1}{\abs{\mathbb{T}}} \sum_{\bfR \in \mathbb{T}} \left[1-F(\bfR)\right]^2,
\end{eqnarray}
where $F(\bfR) = \bra{\psi(\bfR)}\rho^{\rm em}(\bfR)\ket{\psi(\bfR)}$, the error-mitigated state is $\rho^{\rm em}(\bfR) = \sum_{\bfP} q(\bfP) \rho(\bfR,\bfP)$, and $\rho(\bfR,\bfP)$ is the actual noisy final state of the circuit with gate sequences $\bfR$ and $\bfP$. We remark that $F(\bfR)$ is a pseudo fidelity, because $\rho^{\rm em}(\bfR)$ may not be positive. The training circuit $\bfR\in\mathbb{C}$ is Clifford, therefore $\ket{\psi(\bfR)}$ is a stabiliser state~\cite{Gottesman1998}. Suppose $S_{\bfR}$ is the stabiliser group of the state $\ket{\psi(\bfR)}$, we have (see Appendix~\ref{app:fidelity})
\begin{eqnarray}
\bra{\psi(\bfR)}\rho(\bfR,\bfP)\ket{\psi(\bfR)} = \frac{1}{2^n} \sum_{g\in S_{\bfR}} \Tr\left[g\rho(\bfR,\bfP)\right].
\end{eqnarray}
By measuring the group elements $g$, which are Pauli operators with $\pm$ signs, we can evaluate the fidelity and then the loss function. Compared with the loss of one observable, the fidelity loss has an additional summation over the stabiliser group, which can be realised using the Monte Carlo method. 

To measure the operators $g$, usually we need to change the measurement basis. Given the physical measurement setup in the $Z$ basis, we can effectively change the basis by adding single-qubit Clifford gates before the measurement, i.e.~another layer of computing gates. We remark that single-qubit gates next to each other in the circuit can be combined into one single-qubit gate in the physical implementation. Therefore, an additional layer of computing gates does not increase the physical complexity. 

\section{Pauli error model}
\label{sec:Pauli}

In this section, we discuss the Pauli error model, which is the underlying picture of the protocol. By using the error-mitigating gates, we can convert general errors into Pauli errors. In the Pauli twirling method, stochastic Pauli gates are implemented before and after a Clifford gate. Because the gate is Clifford, two sets of Pauli gates cancel with each other if they are properly chosen. Therefore, the Clifford gate is unchanged if it is error-free, but the noise is symmetrised. In Eq.~(\ref{eq:com}), we have Pauli gates before and after the frame-operation tensor, i.e.~$\calP_{\rm L}\calF\calP_{\rm P}$, which is similar to the setup of Pauli twirling of a Clifford gate. Note that $\calF^{\rm ef}$ is a tensor product of Clifford gates except $\calG_0^{\rm ef}$. Errors in the qubit initialisation and measurement, i.e.~$\calG_0$, can also be converted into Pauli errors. See Appendix~\ref{app:Pauli} for details. 

In the following, we assume that errors are Pauli for simplification. We use $[\sigma_1]$ to denote the initialisation error, which occurs after the qubit initialisation, we use $[\sigma_{2j+1}]$ to denote the error of the $j$-th layer frame gate, which occurs after the corresponding frame gate $G_j$, and we use $[\sigma_{2N+2}]$ to denote the measurement error, which occurs before the measurement. Here, $\sigma_j$ are $n$-qubit Pauli operators, $\sigma_1,\sigma_{2N+2}\in\{I,X\}^{\otimes n}$ and $\sigma_{3},\ldots,\sigma_{2N+1}\in\{I,X,Y,Z\}^{\otimes n}$. Referring to Fig.~\ref{fig:circuits}(b) and its obvious generalisation to deeper circuits, we can understand $[\sigma_j]$ as the $j$-th layer of Pauli gates describing errors.

We can use $\bfsigma = \sigma_1\otimes\sigma_3\otimes\cdots\otimes\sigma_{2N+1}\otimes\sigma_{2N+2}$ to describe the pattern of Pauli errors distributed in space-time. If the probability of $\bfsigma$ is $p([\bfsigma])$, the error model can be written as a map $\calN = \sum_{\bfsigma} p(\bfsigma) [\bfsigma]$. Usually, there is an inverse map of $\calN$, which can be written as $\calN^{-1} = \sum_{\bfsigma} q(\bfsigma) [\bfsigma]$, where $q(\bfsigma)$ is the quasi-probability. The distribution $q(\bfsigma)$ is a solution of the equation $\calF^{\rm em} = \calF^{\rm ef}$ and, therefore, can correct all errors for all $\bfR$. With the quasi-probability $q(\bfsigma)$, we take the $j$-th layer of error-mitigating gates as $[\sigma_j]$, where $j=1,3,\ldots,2N+1,2N+2$; error-mitigating gates in other layers are set to identity. 

We can observe that, if the error model is Pauli, error-mitigating gates in $j=2,4,\ldots,2N$ layers are not in fact needed. These layers are only used for general Pauli twirling. 

\section{Significant-error approach}
\label{sec:SigE}

The number of terms in the inverse map $\calN^{-1} = \sum_{\bfsigma} q(\bfsigma) [\bfsigma]$ increases exponentially with the circuit size, and so naively we would require an optimisation of an exponentially many quasi-probabilities $q(\bfsigma)$, which is impractical. In this section and the next three we describe three approaches to practically implement our protocol and provide convincing numerical and quantum hardware experiments with various error models, circuits and tasks. 

As mentioned in Section \ref{sec:practical}, one approach is to assume a Pauli error model $\calN \approx \sum_{\bfsigma\in{\rm SigE}} p(\bfsigma) [\bfsigma]$, where ${\rm SigE}$ as the set of significant errors including the trivial error (i.e.~identity operator). Probabilities of other errors are negligible. If $p(\bfsigma) \ll 1$ for all nontrivial errors, the inverse map is approximately $\calN^{-1} \approx \sum_{\bfsigma\in{\rm SigE}} q(\bfsigma) [\bfsigma]$, which is used as the ansatz in the learning process. This leaves us with a truncated set of optimisation parameters $q(\bfsigma)$ and, by choosing an appropriate construction of the set SigE, it may be truncated to a degree where $q(\bfsigma)$ scales polynomially with the circuit size. 

An example construction of a polynomially scaling set SigE, which we have used in our numerical simulations, is as follows: 

1. Use gate set tomography to find the naive initialisation, measurement and two-qubit gate errors, calculate the respective quasi-probabilities $q(\bfsigma))_{\rm ini}$ for all $\bfsigma$ assuming the error model for the whole circuit is only composed from the combinations of these Pauli errors. Here by `two-qubit gate errors' we mean the error model inferred by an experimentalist purely from tomography of the two-qubit gate mechanism operating on two otherwise-isolated qubits. This task is tractable but will fail to capture the spatial (e.g. cross talk to other qubits) and temporal correlations that will generally occur in the real, comprehensive noise model. Our learning procedure will then adapt the mitigation to encompass these more complex errors. Because the quasi-probabilities eventually used in the error mitigation are determined in the learning, a highly accurate gate set tomography for this initialisation is not required. 

In our numerical simulations, we assume the gate set tomography is accurate, up to the neglected time dependence and correlations, in order to be compared with the learning-based approach. According to the quasi-probability decomposition, the error-correcting gate set is $\{ \bfsigma| q(\bfsigma)_{\rm ini}\neq 0\}$. This set, however, still scales exponentially with the circuit size. This step draws parallels with the protocol introduced in the original probabilistic error cancellation works~\cite{Temme2017, Endo2018}. 

2. To restrict ourselves to a polynomially scaling set SigE, we truncate the error-correcting gate set by leaving only errors up to a constant order $k$, i.e. in any given instance $\bfsigma$ there will be error-mitigating gates $(P_1, P_2, ...)$ associated with at most $k$ of the two-qubit gates. A straightforward extension would be to encompass the initialisation and measurement phases too in order to adapt to correlated errors occurring there, but for our numerical simulations we focus on noise associated with the two-qubit operations.

Similarly, the loss function can be estimated by truncating the complete training set $\mathbb{C}$. We numerically show that the randomly selected subset (i.e.~truncated training set) $\mathbb{T} \subseteq \mathbb{C}$ to a size which is comparable to $c|{\rm SigE}|$ for some overhead constant $c$ is adequate for the learning process. 

After the truncations, the loss function becomes 
\begin{equation}
{\rm Loss} = \frac{1}{|\mathbb{T}|}\sum_{\bfR \in \mathbb{T}}|{\rm com}^{\rm ef}(\bfR,\bfI) - \sum_{\bfsigma\in{\rm SigE}} q(\bfsigma) {\rm com}(\bfR,\bfsigma)|^2.
\end{equation}
Since the sizes of $\mathbb{T}$ and ${\rm SigE}$ scale polynomially with the circuit size, we may evaluate ${\rm com}^{\rm ef}(\bfR,\bfI)$ $\forall \bfR \in \mathbb{T}$ and ${\rm com}(\bfR,\bfsigma)$ $\forall \bfR \in \mathbb{T}$, $\forall \bfsigma \in {\rm SigE}$ using classical and quantum hardware, respectively. Finally, we optimise the truncated quasi-probability $q(\bfsigma)$ using the method of least squares (see Appendix~\ref{app:SigE}). Error-mitigated computation with any circuit $\bfR$ is implemented using $q_{\rm opt}(\bfsigma)$ and the error-mitigation overhead cost $C = \sum_{\bfsigma} \abs{q(\bfsigma)}$. As previously mentioned, this can be implemented either by estimating each ${\rm com}(\bfR,\bfsigma)$ $\forall \bfsigma \in {\rm SigE}$ or by the Monte Carlo summation over SigE. 

Alternative ways to parameterise the quasi-probability distribution will be discussed later.

\subsection*{Numerical simulations}

\begin{figure*}[tbp]
\centering
\includegraphics[width=1\linewidth]{\figpath 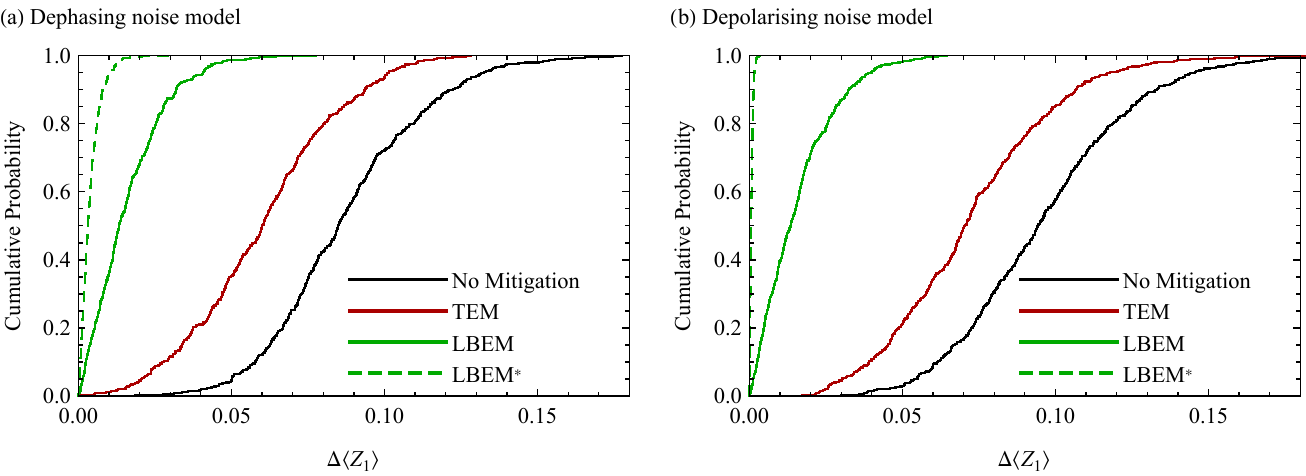}
\caption{
Empirical cumulative distribution function of estimated $\Delta\mean{Z_1}$ for 500 pseudo-random circuits with spatially correlated dephasing noise (a) and spatially correlated depolarising noise (b). Results for circuits without error mitigation (black), with tomographic error mitigation (red) and with learning-based error mitigation (green) are presented. Additionally, we include the results for learning-based error mitigation when sample size $M \to \infty $ (dashed green). 
}
\label{fig:numerics1}
\end{figure*}

\begin{figure}[tbp]
\centering
\includegraphics[width=1\linewidth]{\figpath 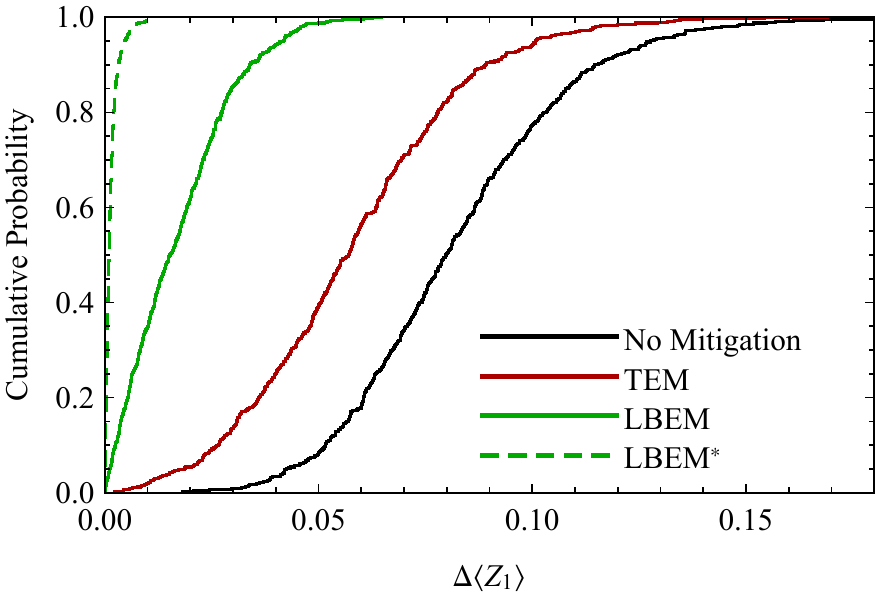}
\caption{
Empirical cumulative distribution function of estimated $\Delta\mean{Z_1}$ for 500 pseudo-random circuits with temporally correlated dephasing noise. Results for circuits without error mitigation (black), with tomographic error mitigation (red) and with learning-based error mitigation (green) are presented. Additionally, we include the results for learning-based error mitigation when sample size $M \to \infty $ (dashed green). 
}
\label{fig:numerics2}
\end{figure}

We present demonstrations of the learning-based quantum error mitigation using the significant-error approach discussed above. We use exact classical simulations of quantum computers with 8 qubits and certain practically-motivated correlated error models. Our simulations are performed using QuESTlink - a Mathematica library which integrates the framework of Quantum Exact Simulation Toolkit (QuEST)~\cite{Jones2020, Jones2019}. The circuits are $n=8$ qubits wide and $N=8$ layers deep for a total of 100 gates in a pattern following Fig. \ref{fig:circuits}(a), where all two-qubit gates are controlled-NOT gates. See Appendix~\ref{app:circuitL} for the detailed circuit. 

We test our error-mitigation scheme with two distinct correlated Pauli error models, one representing spatially and the other temporally correlated noise. In both of these models the local noise, i.e. the noise afflicting the two qubits that are nominally involved in the gate, is homogeneous (dephasing or depolarising) and is assumed to be fully characterised by the experimentalist (either by gate set tomography or pre-existing knowledge). No such assumption is made for the correlated part of the error model. For detailed model please refer to Appendix~\ref{app:EM}.

The set of significant errors SigE is generated from the knowledge of the local noise model and truncated to the $k=1$ order ($|{\rm SigE}| = 85$ or $421$ for dephasing or depolarising noise model respectively). In the loss function we use the deviation from the ideal expectation value of the observable $Z= {\rm diag}(1, -1)$ on the first qubit. The distribution $q_{\rm opt}(\bfsigma)$ is found as indicated above in this section using $|\mathbb{T}|=3|{\rm SigE}|$ filtered randomly generated Clifford circuits (see Appendix~\ref{app:Clif}), where we have chosen Clifford overhead constant $c = 3$ (see Appendix~\ref{app:cScale} for our rationale). 

For a full assessment of the approach, we generate $500$ pseudo-random circuits that satisfy $|\mean{Z_1}^{\rm ef}| > 0.3$ to represent a variety of computational tasks. The restriction to cases with substantial $|\mean{Z_1}^{\rm ef}|$ focuses us on cases where noise can be fully impactful; typically the effect of noise without mitigation is to decrease expected values and thus if a randomly generated circuit happens to produce an expected value close to zero even with zero-noise, then the impact of noise will be minimal. This would obfuscate the performance difference between schemes that provide good mitigation and those that do not. 

Each circuit is formed by drawing its single-qubit computing gates randomly from a circular unitary ensemble. Having performed the learning-based error mitigation once, we apply the same optimised solution to all 500 circuit instances. 
For direct comparison to earlier work, we execute each circuit $M=10000$ times, selecting an appropriate $\bfsigma\in{\rm SigE}$ probabilistically and simply recording a $+1$ or $-1$ for the observable $Z_1$ in each case (inverted if the sign of $q(\bfsigma)$ is negative). In this way we obtain $\mean{Z_1}^{\rm em}$ as a fairly sampled instance of the value that an experimentalist estimates after $M$ samples. We record the absolute deviation 
\[
\Delta\mean{Z_1} = 
|\mean{Z_1}^{\rm em} - \mean{Z_1}^{\rm ef}|
\]
for that circuit, and repeat the process for alternative strategies (tomographic mitigation and no mitigation), before moving to the next of the $500$ circuits. The results are displayed in Fig.~\ref{fig:numerics1}, \ref{fig:numerics2}. 

In the figure, the label `tomography-based error mitigation' refers to the case where the experimentalist has knowledge only of the local error model (i.e. the errors that directly afflict the two qubits nominally involved in a gate) and she samples according to $q_{\rm }(\bfsigma)_{\rm ini}$ $\forall \bfsigma \in {\rm SigE}$ generated with $k=2$ and with the same sample size $M$.

The results presented here are for multi-parameter learning, i.e the elements $q(\bfsigma)$ are independently adjusted during the learning process. In Appendix \ref{app:SinglePO} we include results where the optimisation of $q_{\rm }(\bfsigma)$ is constrained to a single adjustable parameter $\epsilon$, which describes the severity of the local noise. 
$q_{\rm opt}(\bfsigma)$ is then completely defined just by $\epsilon_{\rm opt}$. Note that $\epsilon_{\rm opt}$ is not necessarily equal to the severity of the local noise found from two-qubit tomography.
From the results we can see that such an optimisation strategy yields no better results than tomography-based error mitigation with $q_{\rm }(\bfsigma)_{\rm ini}$ generated with $k=2$, which is slightly above its lower bound on performance set by tomography-based error mitigation with $q_{\rm }(\bfsigma)_{\rm ini}$ generated with $k=1$. However, for sufficiently random circuits and observables, we can expect its performance to increase beyond that of tomography-based error mitigation.

To further test our protocol, we apply it to a hardware efficient variational circuit presented in Appendix~\ref{app:circuitLNew}. The circuit has 8 qubits and consists of 8 layers of random single-qubit rotations around $y$ axis of the Bloch sphere and two-qubit controlled-Z gates and we wish to extract an expectation value of $\sigma_Z$ observable on the bottom qubit, which we denote $\mean{Z_1}$. Qubits are assumed to be laid out in a cycle graph pattern such that a local two-qubit gate may be applied between qubit $i$ and $i+1$ ${\rm mod}$ $n$ or $i-1$ ${\rm mod}$ $n$.

To this circuit we introduce an error model which closer mimics the errors of current NISQ devices compared to the previous error model - single-qubit gates are considered error free compared to two-qubit gates, but are followed by a small probability of relaxation $\gamma$ while two-qubit gates are followed by an error channel

\begin{equation}
\label{eq:channel}
\calD(\epsilon) = (1-\epsilon)[\openone] + \epsilon(\frac{\eta}{\eta+1}\calD^*_{\rm Ph}+\frac{1}{\eta+1}\calD^*_{\rm Pol}),
\end{equation}

where

\begin{equation*}
\calD^*_{\rm Ph} = \frac{1}{3}\sum_{\mu\in \{I,Z\}^{\otimes 2}\setminus I^{\otimes 2}} [\mu],
\end{equation*}

\begin{equation*}
\calD^*_{\rm Pol} = \frac{1}{15}\sum_{\mu\in \{I,X,Y,Z\}^{\otimes 2}\setminus I^{\otimes 2}} [\mu].
\end{equation*}

Here $\eta$ is noise bias between reduced dephasing $\calD^*_{\rm Ph}$ and depolarising channels $\calD^*_{\rm Pol}$ with $\eta=0$ describing a fully depolarising channel and $\eta=\infty$ describing a fully dephasing channel. In our simulations we use $\eta=10$, $\epsilon = 0.01$ and $\gamma = 0.001$. 

\begin{figure}[tbp]
\centering
\includegraphics[scale = 0.8]{\figpath 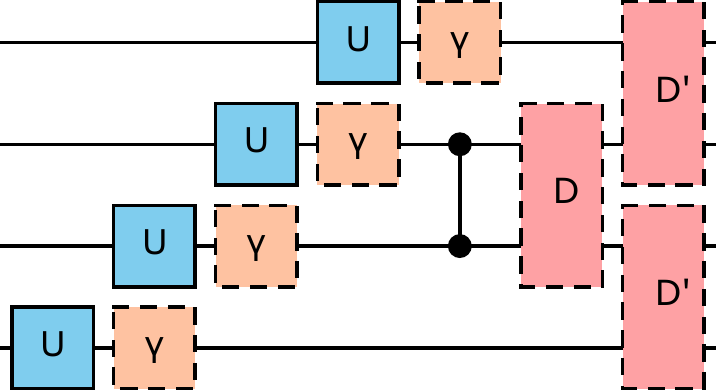}
\caption{
Error model after each set of single qubit gates (blue) and two-qubit gates (black).
Unitary single-qubit gates $U$ may be either single-qubit Clifford gates or arbitrary rotations around $y$ axis of the Bloch sphere. $\gamma$ (orange) describes the amplitude damping channel, while $\calD$ (red) describes a biased dephasing and depolarising channel (the channel is described in the main text, Eq.~\ref{eq:channel}).
}
\label{fig:ErrorCircuit}
\end{figure}

Our protocol is particularly powerful with dealing with correlated noise. To that extent, similarly to the previous numerical study, we introduce additional cross-talk errors that are often unnoticed in local tomographic noise characterisation processes. We simulate these errors by an error channel $\calD' = \calD(\frac{\epsilon}{10})$ which occurs after each two-qubit gate (and its respective error channel described above) between each qubit that is involved and a qubit that is not involved in the two qubit gate, but is locally connected (see Fig.~\ref{fig:ErrorCircuit} to see full error cycle after each controlled-Z gate, for completeness we also show errors after every single-qubit gate layer).

\begin{figure}[tbp]
\centering
\includegraphics[width=1\linewidth]{\figpath 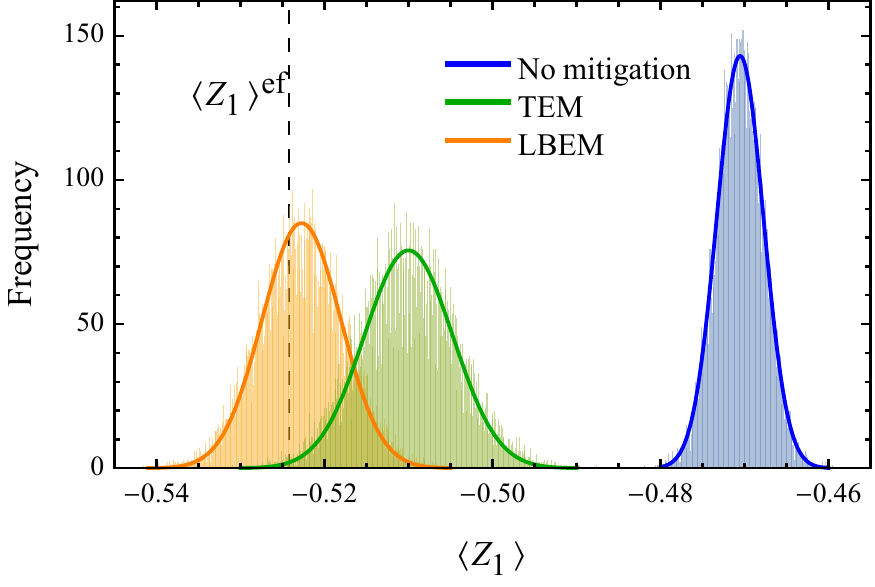}
\caption{
Expectation values of $\mean{Z_1}$ obtained from a single experiment repeated 10000 times with no error mitigation (blue), tomography-based error mitigation (TEM, green) and learning-based error mitigation (LBEM, orange). Dashed line indicates an error-free expectation value $\mean{Z_1}^{\rm ef}$. The solid lines describe analytically derived probability distributions for each approach.}
\label{fig:Result1}
\end{figure}

With this error model we generate a single 8 qubit noisy circuit (Appendix~\ref{app:circuitLNew}) which satisfy $|\mean{Z_1}^{\rm ef}| > 0.5$ to better quantify the effect of our error mitigation protocol. We perform the learning part of the protocol to find to $q_{\rm }(\bfsigma)_{\rm opt}$ $\forall \bfsigma \in {\rm SigE}$ using the significant-error approach with ${\rm SigE}$ being the set of Pauli two-qubit gates after each controlled-Z gate truncated to $k=1$ order. We compare $10^4$ error mitigated expectation values of this circuit to that of an expectation value $\mean{Z_1}^{\rm ef}$ from an error-free circuit after sampling $M= 10^6$ shots according to $|q_{\rm }(\bfsigma)_{\rm opt}|$ for each estimation of the expectation value Fig.~\ref{fig:Result1}. For direct comparison we include non-mitigated expectation values as well as expectation values from previous work on probabilistic error cancellation. In the figure, similarly to the previous numerical study, the label `tomography-based error mitigation' (TEM) refers to the case where the experimentalist has knowledge only of the local error model and she samples according to $q_{\rm }(\bfsigma)_{\rm ini}$ $\forall \bfsigma \in {\rm SigE}$ generated with $k=2$ and with the same sample size $M$. Due to the learning set being truncated to the $k=1$ order and due to the circuit involving non-Pauli error processes (relaxation gates), our protocol does not perfectly mitigate the error, but has substantial improvement compared to the previously studied tomography based error mitigation. Notice that the variance of the LBEM approach is less than the variance of the TEM approach, while achieving closer expectation values to the ideal value. This is because during the learning process, the algorithm finds the required result dependencies on gate errors, and if some error does not affect the computational result outcome (in this circuit some two-qubit gate errors do not affect the resulting expectation value on the bottom qubit), then this error is not corrected to reduce the total quasi-probability overhead and hence the variance. This is another feature of our protocol.


\subsection*{Performance with a variational quantum algorithm}
\label{sec:performance}

\begin{figure}[tbp]
\centering
\includegraphics[width=1\linewidth]{\figpath 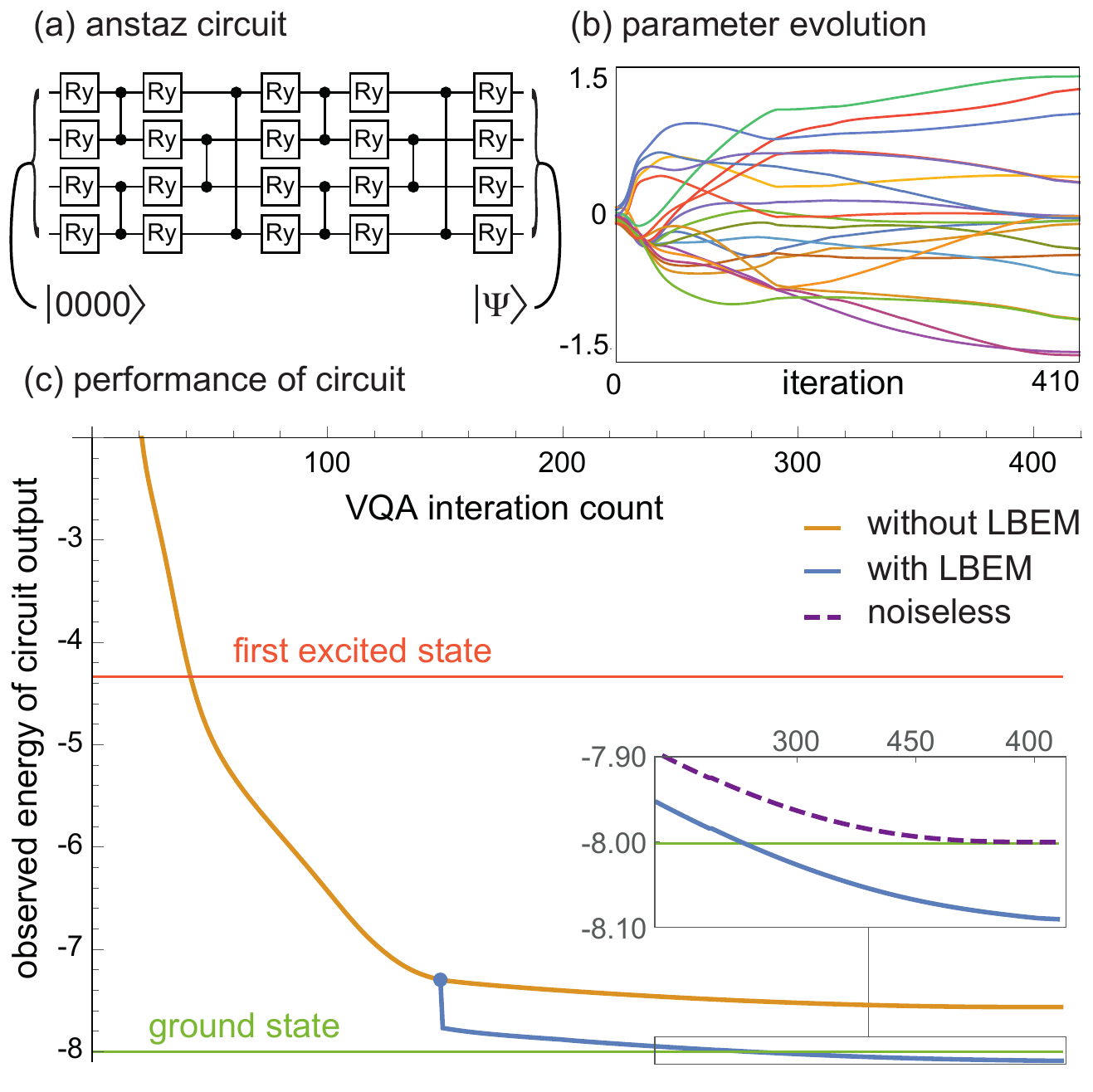}
\caption{
Performance of the learning-based mitigation protocol in the context of a quantum variational algorithm (QVA). The ansatz circuit (a) includes $20$ parameterised gates $R_y(\theta_i)=\exp(-i\theta_i/2 \sigma_y)$ and the parameters are adjusted with the
goal of finding the ground state of a certain frustrated spin system. The processor is a virtual noisy four-qubit device, emulated by the Quantum Exact Simulation Toolkit (QuEST). When the QVA employs simple extrapolation-based mitigation ((c), orange line) the final energy is above the ideal target by $5.6\%$. Instead using learning-based error mitigation, the final energy is below the target by only $0.7\%$. The parameter evolution for the latter case is shown in panel (b). Further details of the two protocols are provided in the main text. 
}
\label{fig:QVA}
\end{figure}

In order explore the efficacy of learning-based mitigation using significant-error approach in a realistic setting, we employed it in the context of a quantum variational algorithm (QVA). The goal of our QVA is to find the ground state energy of a closed chain of four nearest-neighbour interacting spins specified by 
\[
H=\sum_{i=0}^3 A_i \sigma_i^x + J\sum_{i=0}^3\sum_{p\in\{x,y,z\}} \sigma_i^p\sigma_{(i+1)\,\text{mod}\,4}^p
\]
with spins labelled $0$ to $3$. Here the $\sigma$ are the Pauli matrices. We chose $J=1$ and randomly selected the $A$ values. For the data presented in Fig.~\ref{fig:QVA} we used $A =\{0.270777 ,0.192014 ,0.0802803 ,0.123018 \}$), however other simulations had very similar results. This class of system is believed to be classically hard to simulate as the system size grows~\cite{Childs2018}. 

We used a four qubit `ansatz circuit' within which there are $28$ gates: $8$ two-qubit phase gates and $20$ single-qubit rotations (see Fig.~\ref{fig:QVA}). Each single-qubit gate was associated with a unique classical parameter (the rotation angle) and the VQA proceeded by adjusting these parameters in order to  minimise the expected energy $\langle H\rangle$ of the output state, which is therefore the task's cost function. The optimisation method was a canonical gradient descent using the `parameter shift' method~\cite{Schuld2019, Koczor2020} to estimate the gradient with respect to each parameter. 

Note that while the circuit noise severity and system size in this task are consistent with currently available quantum hardware `in the cloud', the very large number of circuit executions required for QVA execution make it cost-prohibitive to use such a device in this context; instead we employed the QuESTlink emulation environment which, as mentioned earlier, has comprehensive and exact noise modelling capabilities. 

The noise model here is similar to the previous one used in above numerical analysis but instead uses higher severity errors with $\epsilon = 0.04$ and $\gamma = 0.002$ (see Eq.~\ref{eq:channel} and Fig.~\ref{fig:ErrorCircuit}). Noise severity was increased in order to achieve a higher contrast between the mitigation schemes given QVA's remarkably high resistance against general noise (as explored in e.g.~\cite{Sharma2020}). We execute our learning-based QEM algorithm by optimising quasi-probability distributions of a significant-error ansatz for each term in the Hamiltonian separately. In our learning process $\rm{SigE}$ is truncated to the $k=1$ order and we take $c=3$.

We would expect that the QVA with the use of learning-based mitigation would far surpass the performance of the same process without any mitigation; therefore for a more meaningful appraisal we compare the learning-based method with the most commonly used alternative mitigation method, i.e. `extrapolation'. In this approach, the desired observables are evaluated both with the lowest possible error rates and with an intentionally boosted error rate, so to estimate the impact of noise and thus to extrapolate to the zero-noise limit. For the present case we assume that the dominant noise type, i.e. the biased mixedness increasing channel, is fully controllable by the experimentalist in the sense that it can be increased to any level with perfect accuracy. However the minor {\it correlated} noise contribution is not under the experimentalist's control in this fashion, and is instead fixed. 

The orange line in panel (c) of Fig.~\ref{fig:QVA} shows how the QVA performs when extrapolation-based mitigation is replied upon. The method works reasonably well considering the very high noise burden; the expected value of the output energy falls from an initial $+4.58$ to $-7.57$ whereas the true ground state energy of the target systems is $-8.002$. Thus the extrapolation method has an absolute energy defect of $3.4\%$ of the spectral width. The blue line indicates the performance when learning-based mitigation is activated at the point when the extrapolation method becomes slowly-evolving. The abrupt downward shift is due to the change in the means of evaluating the energy, i.e. even without changing the ansatz parameters we immediately gain advantage from switching the energy estimation method. There is then a further period of optimisation;  ultimately the energy estimate drops slightly {\it below} the true ground state to $-8.09$, so that the absolute defect is $0.71\%$ of the spectral range. The performance in both methods is in the limit of high sampling, i.e. we presume that the experimentalist is willing to dedicate sufficient repetitions to the process to achieve these optimal trajectories. 

It is notable that although the dominant noise component can be perfectly adjusted for extrapolation (an idealisation that favours that technique), and the non-adjustable component is an order of magnitude smaller, nevertheless the ultimate output of the QVA when using learning-based mitigation is nearly five times superior to the extrapolation protocol (achieving a defect of only $0.71\%$ rather than $3.4\%$). 

\section{Product-form ansatz approach}
\label{sec:product-form}



\begin{figure*}[tbp]
\begin{center}
\includegraphics[width=1\linewidth]{\figpath/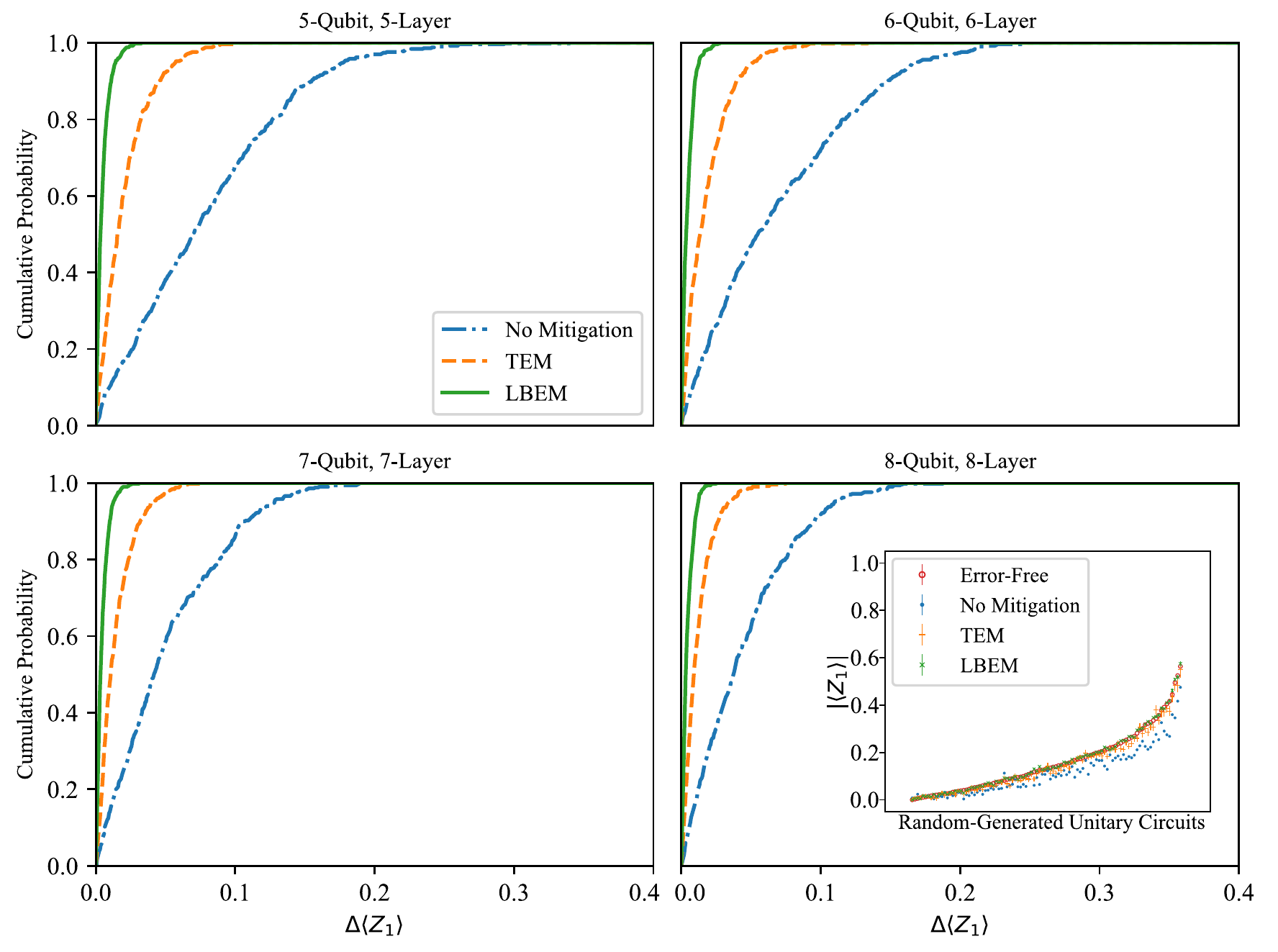}
\caption{
Empirical cumulative distribution function of estimated $\Delta \langle Z_{1}\rangle = \abs{ \langle Z_{1}\rangle - \langle Z_{1}\rangle^{\rm ef} }$ for $500$ configurations of randomly generated computing gates $\bfR$. Each computing gate is uniformly sampled from the single-qubit unitary group according to the Haar measure. For each computing-gate configuration, $M = 10000$ random configurations of error-correcting gates $\bfP$ are generated to evaluate the error mitigated result. TEM and LBEM stand for results with tomography-based error mitigation and learning-based error mitigation, respectively. In the inset in (d), we show $\langle Z_{1}\rangle$ of $100$ configurations for the eight-qubit circuit. Error bars represent estimated standard errors. 
}
\label{fig:unitary_test_combined}
\end{center}
\end{figure*}

Another practical approach to implement the learning based error mitigation protocol is by considering an error ansatz whose distribution admits a product-form described below.



We denote the quasi-probability of each Pauli gate $\bfP_i\in {\rm SigE}$ as $q_i$, and we shall optimise $q_i$ in the learning process. According to the product-form ansatz, we have 
\begin{eqnarray}
{\rm com}^{\rm em}(\bfR,\bfI) = \sum_{\bfb}  q_{\bfb} {\rm com}(\bfR,\bfP_{\bfb}),
\end{eqnarray}
where $\bfb = (b_1,b_2,\dots,b_{\abs{\rm SigE}})$ is a binary vector, and $b_i = 0,1$ denotes that the $i$-th Pauli gate $\bfP_i\in {\rm SigE}$ is off or on. Here, 
\begin{eqnarray}
q_{\bfb} = \prod_{i=1}^{\abs{\rm SigE}} \left[ b_iq_i + (1-b_i)(1-q_i) \right]
\end{eqnarray}
is the quasi-probability distribution of the Pauli gate configuration 
\begin{eqnarray}
\bfP_{\bfb} = \prod_{i=1}^{\abs{\rm SigE}} \bfP_i^{b_i}.
\end{eqnarray}
We note that we have used $\bfP\bfP' = (P_1P'_1,P_2P'_2,P_3P'_3,\ldots)$ to denote the product of two Pauli strings, and $P_jP'_j$ is a Pauli operator up to a phase that can be ignored. In this approach we always have $\sum_{\bfb}  q_{\bfb} = 1$. 

Details for evaluating and minimising the loss function, as well as, learning rates used for different size circuits can be found in the Appendix~\ref{app:LossFn}. 

\subsection*{Numerical simulations}

\begin{table}[tbp]
\begin{tabular}{|c|c|c|c|}
\hline 
Circuit size & $\epsilon'$ & Circuit size & $\epsilon'$\tabularnewline
\hline 
\hline 
$5\times5$--$8\times8$ & 1 & $15\times15$--$16\times16$ & 1.2\tabularnewline
\hline 
$9\times9$--$12\times12$ & 1.1 & $19\times19$--$20\times20$ & 1.3\tabularnewline
\hline 
\end{tabular}
\caption{
Error rate parameter $\epsilon'$ used for various size circuits. Circuit sizes are described in a short-hand notation by $n \times N$ with $n$ qubits and $N$ layers.
}
\label{tableI}
\end{table}

We demonstrate the product-form ansatz approach by numerically simulating various size noisy quantum circuits with the same layout as in Appendix~\ref{app:circuitL}. We take the error model to be spatially correlated depolarising error model, introduced in Section~\ref{sec:SigE}. That is, for each two-qubit gate on qubits $i$ and $i+1$, the error rate of two-qubit depolarising channel on qubits $i$ and $i+1$ is $\epsilon = 2\epsilon'/Nn$ for an $n$-qubit $N$-layer circuit, where $\epsilon'$ is given in Table~\ref{tableI}. To add spatially correlated noise, we also apply two-qubit depolarising channels on qubits $i-1 \mod n$ and $i$ as well as qubits $i+1$ and $i+2 \mod n$ with the error rate $\epsilon/10$.


To show the effect of error mitigation, we test the computation accuracy before and after error mitigation using random unitary single-qubit gates. The results are shown in Fig.~\ref{fig:unitary_test_combined}. Because simulating quantum circuits with general single-qubit unitaries is costly, we only benchmark circuits with the size up to $8$ qubits with $8$ layers. Note that in these simulations we do not require the pseudo-random circuits to satisfy some value of $|\mean{Z_1}^{\rm ef}|$.

\subsection*{Average error rescaling factor}
\begin{figure}[tbp]
\begin{center}
\includegraphics[width=1\linewidth]{\figpath/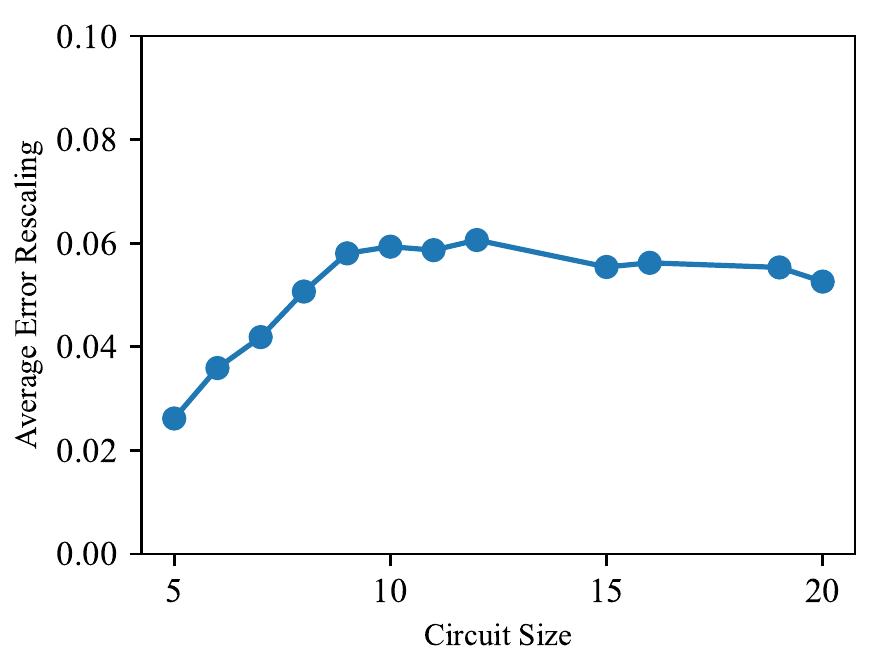}
\caption{
Average error rescaling factor for various circuit sizes. The circuit size is equal to the number of qubits and layers in the circuit. For each circuit layout, we randomly generate $1000$ configurations of Clifford computing gates $\bfR$. We choose the configuration such that the error-free computation result is non-zero. For each computing-gate configuration, $M = 1000000$ random configurations of error-correcting gates $\bfP$ are generated to evaluate the error mitigated result. 
}
\label{fig:rescaling}
\end{center}
\end{figure}

To numerically demonstrate the effect of error mitigation for larger circuits (up to 20 qubits), we test the computation accuracy before and after error mitigation using configurations of Clifford computing gates, such that the circuit can be efficiently simulated using a classical computer. We use the average error rescaling factor to quantify the effect of error mitigation, which is defined as 
\begin{eqnarray}
r = \left\langle\frac{\left\vert {\rm com}^{\rm em}(\bfR,\bfI)-{\rm com}^{\rm ef}(\bfR,\bfI) \right\vert}{\left\vert {\rm com}(\bfR,\bfI)-{\rm com}^{\rm ef}(\bfR,\bfI) \right\vert}\right\rangle.
\end{eqnarray}
The error rescaling factor as a function of the circuit size is plotted in Fig.~\ref{fig:rescaling}. We can find that the error rescaling factor does not increase with the circuit size when the size is larger than $9$ qubits, indicating the efficient scalability of our protocol. The remaining error after error mitigation is mainly due to the statistical fluctuations caused by a finite number of samples in the learning and error mitigation stages. 

\section{Variational distribution and Monte Carlo evaluation}
\label{sec:Boltzmann}

In addition to the summation and product form ansatz of the Pauli error model described above 
we can potentially use variational functions such as the restricted Boltzmann machine to tackle error models with unknown features~\cite{Huo2017}. The restricted Boltzmann machine~\cite{Fischer2012} can efficiently express the distribution in a large state space, which can represent the complex-valued wavefunction by using complex weights~\cite{Carleo2017}. The quasi-probability distribution is real-valued in our case. 

In general, we can express the ansatz in the form $q(\bfP) = CB(\bfP,\lambda)/A(\lambda)$, where $C$ and $\lambda$ are variational parameters that are optimised in the learning process. The function $B(\bfP,\lambda)$ must be computable on the classical computer, and $A(\lambda) = \sum_{\bfP} \abs{B(\bfP,\lambda)}$ is the normalisation factor. Even if we cannot compute $A(\lambda)$, samples of the distribution $\abs{B(\bfP,\lambda)/A(\lambda)}$ can be efficiently generated using the Metropolis method. The number $C = \sum_{\bfP} \abs{q(\bfP)}$ is the error-mitigation overhead cost~\cite{Endo2018}. When we already have the optimal parameters, we can implement the error-mitigated computing by using the Monte Carlo summation with samples of $\bfP$ generated according to the optimal distribution. The variance of the error-mitigated computing is ${\rm Var}\left[\hat{\rm com}^{\rm em}\right] \leq \frac{1}{M} \abs{f}_{\rm max}^2 C^2$, where $M$ is the number of samples, and $\abs{f}_{\rm max}$ is the maximum value of $\abs{f(\bfmu)}$. Here, we have assumed that the circuit only runs for once (without repeating) for each sample of $\bfP$. 

The Monte Carlo method can also be used to compute the loss function. The loss function is in the quadratic form with respect to $C$. Therefore, it is straight-forward to find the optimal $C$ given the value of $\lambda$. To find the optimal $\lambda$, we usually need to evaluate the loss for different values of $\lambda$. Instead of generating samples for each value, we can compute the loss for $\lambda$ using samples generated according to a different value $\lambda'$. In this way, we can reduce the sampling cost in the learning process. Once the optimal $\lambda$ is found, we need to generate samples according to the optimal $\lambda$ in order to compute the optimal $C$. The variance of the loss is ${\rm Var}\left[\hat{\rm Loss}\right] \lesssim \frac{1}{M} \abs{f}_{\rm max}^4(1+C^4+4C^2)$, if samples are generated according to the same value of $\lambda$. 

The details of the Monte Carlo summation, including the application in the significant-error approach, can be found in Appendix~\ref{app:MC}.

\section{Experimental demonstrations}
\label{sec:experiment}

\subsection*{Two-qubit DQCp circuit}

\begin{figure}[tbp]
\centering
\includegraphics[width=1\linewidth]{\figpath/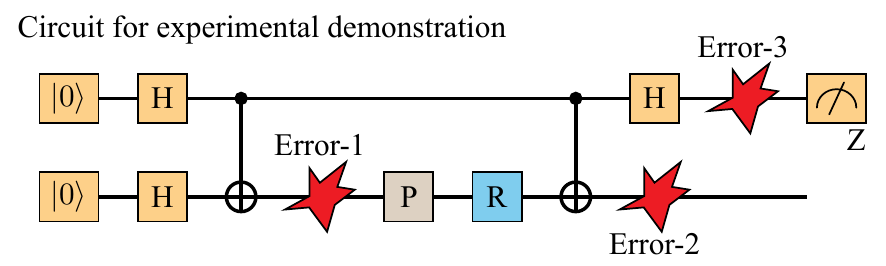}
\caption{
Two-qubit DQCp circuit used in the experimental demonstration of the learning-based quantum error mitigation protocol. 
}
\label{fig:circuit2q}
\end{figure}

\begin{figure*}[tbp]
\centering
\includegraphics[width=1\linewidth]{\figpath/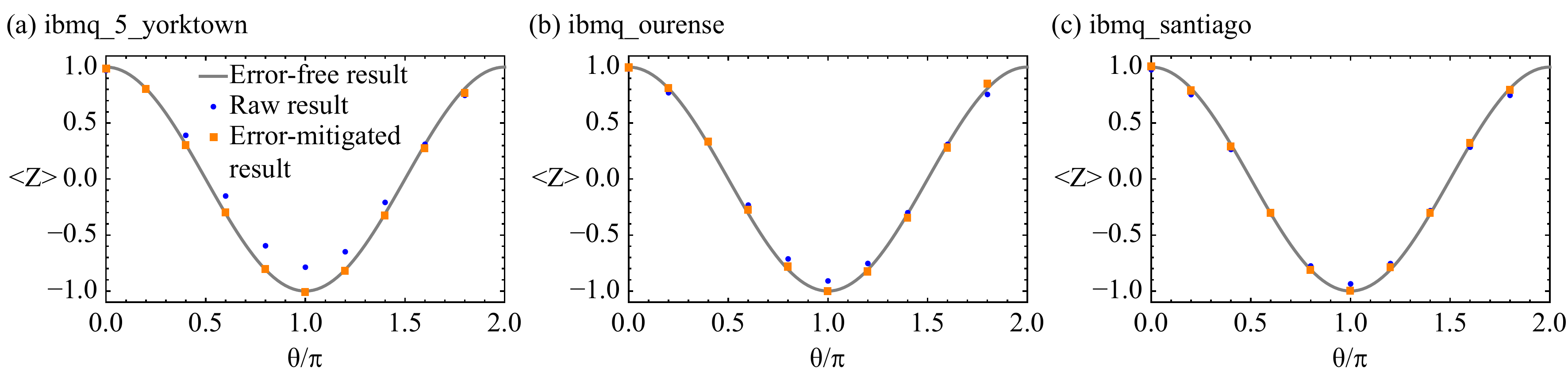}
\caption{
Computation results of the two-qubit DQCp circuit, obtained from three different quantum hardware. $\mean{Z} = {\rm com}(e^{-i\frac{\theta}{2}Z},I)$ is the raw result without error mitigation, and $\mean{Z} = {\rm com}^{\rm em}(e^{-i\frac{\theta}{2}Z},I)$ is the error-mitigated result. 
}
\label{fig:IBMq}
\end{figure*}

We demonstrate our learning-based QEM protocol on three IBMQ machines, \textit{ibmq\_5\_yorktown}, \textit{ibmq\_ourense} and \textit{ibmq\_santiago}. On all three cases we observe an improvement of the computation result when executing a two-qubit DQCp circuit given in Fig.~\ref{fig:circuit2q}. Taking $R=e^{-i\frac{\theta}{2}Z}$ and $P=I$, the error-free result (which we take to be the mean of $Z$ of the upper qubit) is given by ${\rm com}^{\rm ef}(e^{-i\frac{\theta}{2}Z},I)=\cos(\theta)$, as shown in Fig.~\ref{fig:IBMq}.

To perform this demonstration we simplified our protocol to reduce the amount of Pauli gates we introduce in the circuit compared to the original protocol in which layers of Pauli gates are being used, Fig.~\ref{fig:circuits}. This is done by assuming a Pauli error model and noting that all Clifford gates map Pauli errors back to other Pauli errors. Hence, we only need to introduce error correcting Pauli gates before any non-Clifford gate (gate $R$ in Fig.~\ref{fig:circuit2q}). In our circuit only a single Pauli gate is inserted to correct the error labeled $\text{Error-1}$, while errors labeled $\text{Error-2}$ and $\text{Error-3}$ either do not impact the computational result or act as a measurement error. The measurement errors can be corrected by modifying the original formula of error mitigation Eq.~\ref{eq:errorMit}. 

For the two-qubit DQCp circuit, the computation result with the error mitigation can be written as
\begin{eqnarray}
{\rm com}^{\rm em}(R,I) = \sum_{P=I,X,Y,Z} q(P) {\rm com}(R,P) + q_0,
\label{eq:2qQEM}
\end{eqnarray}
where $q_0$ is the term associated with the measurement error. Suppose the measurement error can be modelled as follows: The measurement outcome is flipped with probability $p_{\mu}$ for some correct output state $\ket{\mu}$, where $\mu=0,1$. According to this model, the mean value $\mean{Z}_c$ describing the correct measurement outcome and the mean value $\mean{Z}_e$ describing the erroneous measurement outcome have a simple relation $\mean{Z}_c = (\mean{Z}_e+p_0-p_1)/(1-p_0-p_1)$. Hence, the optimal value of $q_0$ is given as $(p_0-p_1)/(1-p_0-p_1)$. We may include the factor $1/(1-p_0-p_1)$ in the quasi-probability distribution $q(P)$, e.g. if the quasi-probability distribution for correcting $\text{Error-1}$ is $q'(P)$, we have $q(P) = q'(P)/(1-p_0-p_1)$.

In the learning part of the algorithm, we ran $24$ different circuits on each of the three IBMQ quantum machines to evaluate ${\rm com}(C_i,I)$, where $C_i$ is one of the $24$ single-qubit Clifford gates. By minimising the loss function 
\begin{eqnarray}
{\rm Loss} &=& \frac{1}{24} \sum_{i=1}^{24} \Big[ \sum_P q(P) {\rm com}(C_i,P) + q_0 \notag \\
&&- {\rm com}^{\rm em}(C_i,I)\Big]^2
\label{eq:DQC2}
\end{eqnarray}
we obtain optimal quasi-probability distribution $q(P)$ and optimal $q_0$.
We note that ${\rm com}(C_i,P) = {\rm com}(C_iP,I)$, and that $C_iP$ is one of the $24$ single-qubit Clifford gates, therefore all ${\rm com}(C_i,P)$ can be derived from the set $\{{\rm com}(C_i,I)\}$. Since the loss is a quadratic function, the minimisation is straightforward.

Next, we test our error mitigation protocol by taking $R=e^{-i\frac{\theta}{2}Z}$, where $\theta = 2m\pi/10$, and $m=0,1,\ldots,9$. The results are shown in Fig.~\ref{fig:IBMq}. For each machine, we implement $40$ circuits to evaluate ${\rm com}(e^{-i\frac{\theta}{2}Z},P)$ with $\theta$ taking ten different values, and $P=I,X,Y,Z$. We write ${\rm com}(e^{-i\frac{\theta}{2}Z},I)$ to denote computation results without error mitigation, which deviate from the error-free values due to the quantum hardware being noisy. The error-mitigated results are computed according to Eq.~(\ref{eq:2qQEM}), in which we take $R=e^{-i\frac{\theta}{2}Z}$ and the optimal values of $q(P)$ and $q_0$ obtained by minimising the loss function. It is clear that the error mitigation reduces the computation error.

Potential causes of residual errors after error mitigation are statistical fluctuations and non-Pauli errors. For each circuit, we run $8192$ shots to evaluate the mean value $\mean{Z}$. Pauli twirling is not used in this experiment, i.e.~general errors are not converted into Pauli errors. Even then, our error mitigation protocol can significantly improve the computation result accuracy, which demonstrates the robustness of the learning approach. We remark that the error mitigation of two-qubit DQCp circuit has been demonstrated in Ref.~\cite{Song2019}, in which gate set tomography is used to work out the quasi-probability distribution. This tomography of a two-qubit gate requires at least $256$ circuits, while in our approach only $24$ circuits are used in the learning part for determining the error mitigation parameters. 

\subsection*{Variational quantum eigensolver}

\begin{figure}[tbp]
\centering
\includegraphics[width=1\linewidth]{\figpath 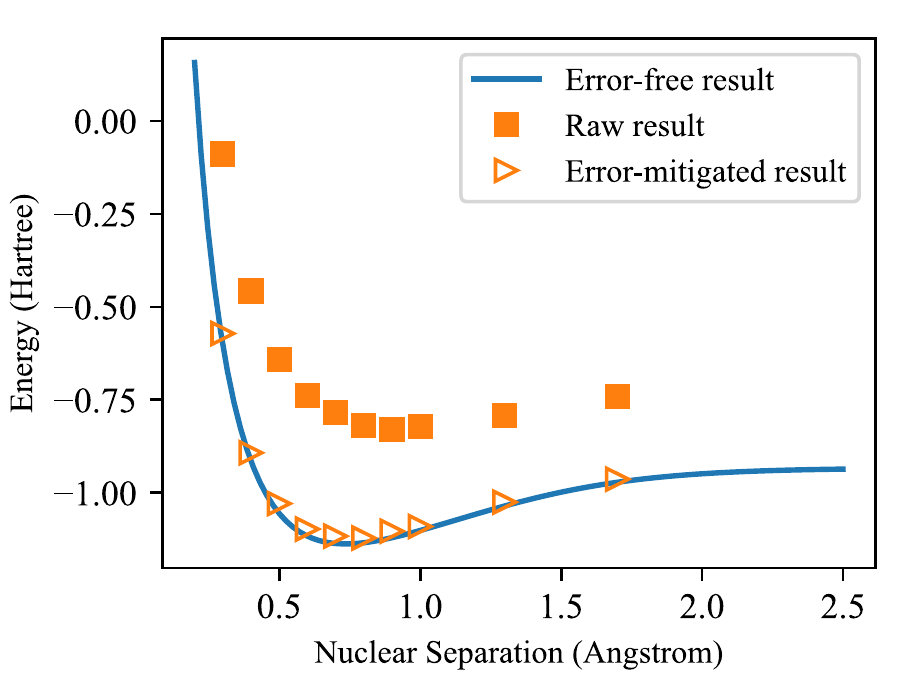}
\caption{
Ground state energy surface of $\text{H}_{\text{2}}$ in the minimal basis computed using variational quantum eigensolver. The blue solid line is computed using package \textit{Qiskit}. Square and triangular scatters represent results without and with the learning-based quantum error mitigation computed on \textit{ibmq\_santiago}. 
}
\label{fig:energy_surface}
\end{figure}

In addition to the two-qubit DQCp circuit, we also experimentally demonstrate our learning-based quantum error mitigation (LBEM) protocol by applying it to the variational quantum eigensolver (VQE) algorithm. We compute the ground state energy of $\text{H}_{\text{2}}$ molecule at different nuclear separations on IBMQ machine \textit{ibmq\_santiago} with and without LBEM. The results are shown in Fig.~\ref{fig:energy_surface}. We find that LBEM can significantly improve the accuracy of VQE. 

In this demonstration, we compute the ground state energy of $\text{H}_{\text{2}}$ in the minimal basis (STO-3G basis), which includes 4 spin-orbitals (each atom contribute two spin-orbitals $\{1s_{\uparrow},1s_{\downarrow}\}$). The electronic wavefunction are projected onto these 4 spin-orbitals, and then we use the Jordan-Wigner transformation to map fermions to qubits. The corresponding Hamiltonian of qubits reads
\begin{equation}
H = H_{1} + H_{2},\label{eq:hjw}
\end{equation}
where 
\begin{align}
H_{1} & =h_{0}I+h_{1}Z_{0}+h_{2}Z_{1}+h_{3}Z_{2}+h_{4}Z_{3}\nonumber \\
 & +h_{5}Z_{1}Z_{0}+h_{6}Z_{2}Z_{0}+h_{7}Z_{3}Z_{0}\nonumber \\
 & +h_{8}Z_{2}Z_{1}+h_{9}Z_{3}Z_{1}+h_{10}Z_{3}Z_{2}\label{eq:h1}
\end{align}
and 
\begin{align}
H_{2} & =h_{11}X_{3}X_{2}Y_{1}Y_{0}+h_{12}Y_{3}Y_{2}X_{1}X_{0}\nonumber \\
 & +h_{13}X_{3}Y_{2}Y_{1}X_{0}+h_{14}Y_{3}X_{2}X_{1}Y_{0}.\label{eq:h2}
\end{align}
Here we have written $H$ into two parts according to the commutation relation between Pauli operators. Pauli operators in $H_{1}$ ($H_{2}$) commute with each other, therefore they can be measured using the same circuit. We use \textit{Qiskit} to calculate the coefficients in Eq.~(\ref{eq:h1}) and Eq.~(\ref{eq:h2}).

\begin{figure*}[tbp]
\centering
\includegraphics[width=1\linewidth]{\figpath 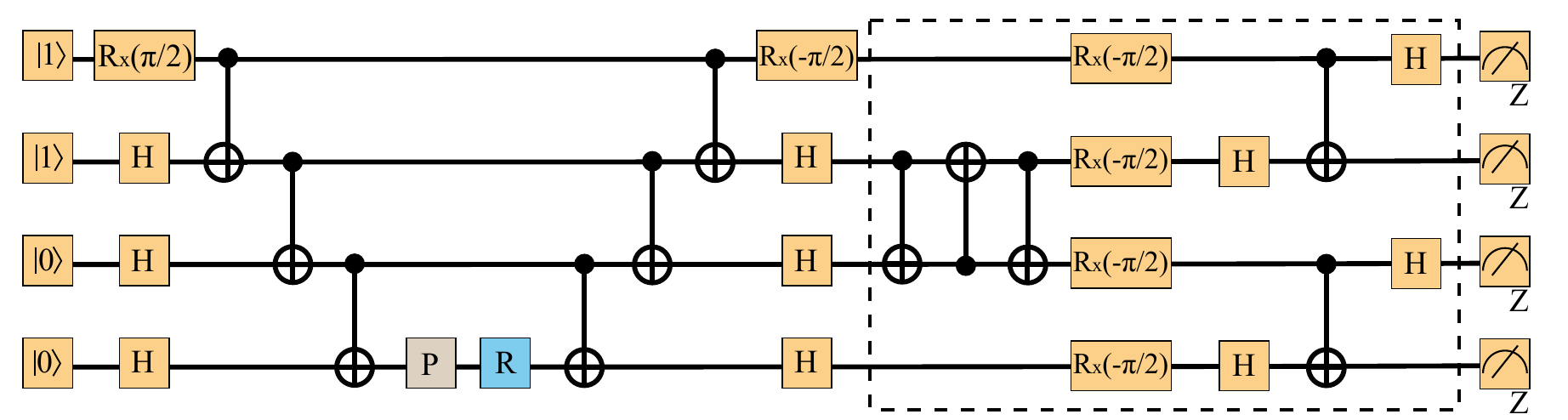}
\caption{
Circuit used in the error-mitigated variational quantum eigensolver. The gates $R_x(\phi) = e^{-i\frac{\phi}{2}X}$. 
}
\label{fig:VQEcircuits}
\end{figure*}

Similar to the two-qubit DQCp circuit, we implement the error-mitigated VQE by randomly inserting a Pauli gate into the UCCSD-circuit (unitary coupled cluster ansatz truncated to single and double excitations). We directly adapt the simplified UCCSD-circuit given in Ref.~\cite{McArdle2020}, shown in Fig.~\ref{fig:VQEcircuits}(a), parameterized by only one rotational angle of a single-qubit gate $R = e^{-i\frac{\theta}{2}Z}$. The Pauli gate $P$ is inserted before the gate $R$. Without gates in the dashed box, the circuit can be used to evaluate the mean of $H_{1}$, while gates in the dashed box effectively change the measurement basis and transform $H_{2}$ into 
\begin{align}
H_{2}^{\prime} & =h_{11}Z_{3}Z_{1}+h_{12}Z_{2}Z_{0}\nonumber \\
 & +h_{13}Z_{3}Z_{0}+h_{14}Z_{2}Z_{1}.\label{eq:h2prime}
\end{align}
Then, we can use the circuit with gates in the dashed box to evaluate the mean of $H_{2}$.

There are ten non-trivial Pauli operators in $H_{1}$ and four Pauli operators in $H_{2}$. We apply LBEM to each Pauli operator individually, i.e. ${\rm com}(R,P)$ is the mean of one Pauli operator, hence, the error mitigation for one Pauli operator is the same as for the two-qubit DQCp circuit - the error-mitigated computation result of the Pauli operator is given by Eq.~(\ref{eq:2qQEM}), and the loss function is in Eq.~(\ref{eq:DQC2}). At the learning stage, similarly $24$ Clifford gates are implemented instead of the $R$ gate for each one of the two circuits, with and without gates in the dashed box in Fig.~\ref{fig:VQEcircuits}. The data are used to obtain coefficients $q(P)$ and $q_0$. We note that the coefficients are different for each Pauli operator. To demonstrate the effect of LBEM, we take ten different $R$ gates, where each of them is the optimal gate in VQE that minimizes the mean of the Hamiltonian for a given nuclear separation, Fig.~\ref{fig:energy_surface}.

\section{Conclusions}
\label{sec:Conclusions}

In this paper we present a novel way of mitigating quantum errors based on probabilistic error cancellation technique. We introduce a new learning component of the protocol which replaces the need of reconstructing an error model in the experiment. The learning component exploits the efficient simulatability of Clifford circuits and finds the optimal quasi-probability distribution which then defines the next step of probabilistic error cancellation. 
Numerically, we have shown that the learning-based protocol can be practically implemented for the circuit sizes comparable to those currently run on NISQ era quantum computers. In the presence of correlated noise, it outperforms the tomography-based protocol for which tomography on a smaller subset of qubits is only available. We confirm that our protocol maintains its high performance on real quantum hardware by running multiple experiments on the IBM quantum devices.

Different tactics may be employed for learning the optimal quasi-probability distribution depending on the quantum device at hand and the required computations. For example, if one wants to evaluate multiple observables of a computation, so does the learning process needs to include these observables. Using fidelity as a cost function is also a valid strategy, which is not difficult to estimate for Clifford circuits, and then error mitigated expectation value can be estimated for any observable. Similarly, procedures for estimating mean values of functions can be specifically tailored, for example, in cases where modifying the circuit between consecutive runs is difficult or expensive.

Possible extensions to this work include specifically modifying the learning component of the protocol in cases where some information about the noise model is given or easily accessible, for example, a scenario where the whole circuit undergoes an unknown global phase shift. Another extension would be to sample Clifford circuits respective to the unitary circuits they replace. This would lead to a circuit specific error mitigation. The same learning approach based on Clifford circuit sampling can also be applied for finding the optimal physical parameters of a quantum computing system.

There are of course a number of quantum error mitigation techniques that have already been proposed in the literature. Each of the these protocols have their advantages and disadvantages; in some cases our method represents an alternative and in other cases the methods can be concatenated. For example: given complete knowledge of the noise processes in a quantum system, theoretically one can compensate perfectly for errors using the quasi-probability error mitigation technique~\cite{Temme2017, Endo2018}. In the previous literature (see~\cite{Endo2018}), the required decomposition formula is worked out using the gate set tomography, which practically yields high performance only when the error correlations are negligible. The present protocol is an alternative where adaptive learning is shown to be capable of replacing the exhaustive tomography, achieving a near-ideal outcome with profoundly reduced cost. 

Another common protocol of quantum error mitigation is the noise extrapolation technique which can efficiently suppress the errors by boosting the noise and extrapolating to the zero limit~\cite{Li2017, Temme2017, Kandala2019}. However, due to the discrepancy between the fitting curve and the genuine curve of ``computation result vs. noise'', and challenge of homogeneously boosting noise, practically speaking extrapolation cannot be expected to perfectly compensate for errors~\cite{Cai2020}. 
Finally, error mitigation based on symmetries post-selection can  correct errors that violate such symmetries; it is a powerful method where such symmetries exist~\cite{McArdle2019, Bonet2018}.

One day when we use quantum computer to solve some meaningful practical problems, we may need to combine different error mitigation protocols in order to achieve the required accuracy. The learning-based approach is flexible and can be used as a framework of error mitigation that serves the purpose of integrating different protocols. For example, the learning-based approach can be applied to noise extrapolation, e.g. use the loss function to determine how to choose the fitting curve and how to boost the noise. We can combine noise extrapolation, post-processing based on symmetries and quasi-probability decomposition techniques, and optimise the overall error mitigation strategy by again using the loss function. In this way, advantages of all the mentioned protocols may be exploited.

Overall, our protocol paves a new way of implementing NISQ era quantum error mitigation and is especially suitable for remote users without any access to the information about the noise model. It is intuitively simple and can be readily implemented on current quantum computers. 

\vspace{6mm}
\textit{Note added.} Shortly after the original version of this work was posted online, a related work was posted by P. Czarnik \textit{et al.}~\cite{Czarnik2020}. They exploit `near-Clifford' circuits to obtain an error-mitigated estimator of some observable for a circuit of interest. Both papers reveal the power of Clifford variants in error mitigation, however the methods differ fundamentally in ways the error mitigation is applied, noise assumptions made and necessity of non-Clifford gates in the learning part of the algorithm.

\begin{acknowledgments}
DYQ and YL are supported by National Natural Science Foundation of China (Grant No. 11875050) and NSAF (Grant No. U1930403). YC acknowledges support from BNL LDRD \#19-002 and the National Science Foundation (Grant No. PHY 1915165). SCB acknowledges support from the EU Flagship project AQTION, the NQIT Hub (EP/M013243/1) and the QCS Hub (EP/T001062/1).
\end{acknowledgments}

\appendix

\section{The formalism of quantum circuits with a frame}
\label{app:formalism}

We consider a circuit with $n$ qubits and $N$ layers of frame gates between the qubit initialisation and measurement. All qubits are initialised in the state $\ket{0}$ at the beginning and measured in the $Z$ basis at the end. Each layer of frame gates is formed by multi-qubit Clifford gates, as shown in Fig.~\ref{fig:circuits}(a). Single-qubit unitary gates are between frame operations (including the qubit initialisation, frame gates and measurement), and we call them computing gates. The frame operations are fixed, but computing gates are treated as variables. For the error mitigation, single-qubit Pauli gates are introduced before and after each computing gate, as shown in Fig.~\ref{fig:circuits}(b). We call these Pauli gates error-mitigating gates, which are also variables. 

\subsection{Notations}

We use $\bfR \equiv (R_1,R_2,\ldots,R_{n(N+1)})$ and $\bfP \equiv (P_1,P_2,\ldots,P_{2n(N+1)})$ to denote the computing gate sequence and error-mitigating gate sequence, respectively. $\bfP = \bfI$ denotes that all error-mitigating gates are identity gates. 

We use $\mu_k = 0,1$ to denote the measurement outcome of the $k$-th qubit. $\bfmu \equiv (\mu_1,\mu_2,\ldots,\mu_n)$ is the binary vector that represents the outcome of all qubits. The task is to compute the mean value of a function $f(\bfmu)$. 

We use ${\rm com}(\bfR,\bfP)$ to denote the mean value of the function $f(\bfmu)$ given the gate sequences $\bfR$ and $\bfP$. ${\rm com}^{\rm ef}(\bfR,\bfP)$ is the value of ${\rm com}(\bfR,\bfP)$ when the entire computing is error-free. 

We use $q(\bfP)$ to denote a quasi-probability function, and the error-mitigated computing result is 
\begin{eqnarray}
{\rm com}^{\rm em}(\bfR,\bfI) \equiv \sum_{\bfP} q(\bfP) {\rm com}(\bfR,\bfP).
\end{eqnarray}
The error function is 
\begin{eqnarray}
{\rm Error}(\bfR) \equiv \absLR{ {\rm com}^{\rm em}(\bfR,\bfI) - {\rm com}^{\rm ef}(\bfR,\bfI) }.
\end{eqnarray}

The loss function of the computing error is 
\begin{eqnarray}
{\rm Loss} \equiv \frac{1}{\abs{\mathbb{T}}} \sum_{\bfR \in \mathbb{T}} {\rm Error}(\bfR)^2,
\end{eqnarray}
where $\mathbb{T}$ is a set of computing gate sequences. The training set $\mathbb{T}$ is a subset of Clifford gate sequences, i.e.~$\mathbb{T} \subseteq \mathbb{C} \equiv \{ \bfR \st \text{All }R_j\text{ are Clifford} \}$. We use $\mathbb{U} \equiv \{ \bfR \st \text{All }R_j\text{ are unitary} \}$ to denote the set of unitary gate sequences, then $\mathbb{T} \subseteq \mathbb{C} \subset \mathbb{U}$. 

\subsection{Quantum formalism}

\begin{figure}[tbp]
\centering
\includegraphics[width=1\linewidth]{\figpath 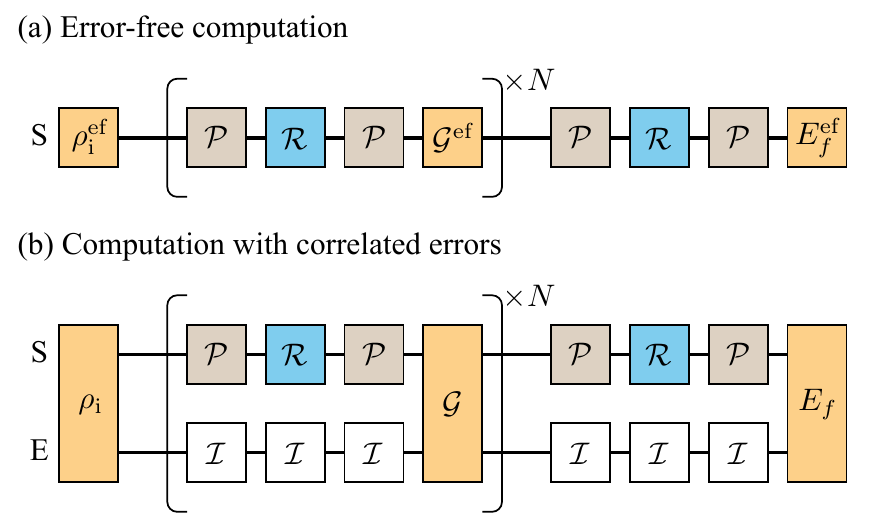}
\caption{
The computing without and with errors. From left to right $\calR = \calR_1,\calR_2,\ldots,\calR_{N+1}$, $\calP = \calP_1,\calP_2,\ldots,\calP_{2N+2}$, $\calG^{\rm ef} = \calG_1^{\rm ef},\calG_2^{\rm ef},\ldots\calG_N^{\rm ef}$ and $\calG = \calG_1,\calG_2,\ldots,\calG_{N}$. $\mathcal{I} = [\openone_{\rm E}]$ is the identity map on the environment. S: system; E: environment. 
}
\label{fig:model}
\end{figure}

We use $\rho_{\rm i}^{\rm ef} \equiv \ketbra{0}{0}^{\otimes n}$ to denote the error-free initial state. We use $[U]\bullet \equiv U\bullet U^\dag$ to denote the completely positive map of the unitary operator $U$. If frame gates are error-free, the overall map of the $j$-th-layer frame gates is $\calG_j^{\rm ef} \equiv [G_j]$, where $G_j$ is an $n$-qubit Clifford gate, as shown in Fig.~\ref{fig:circuits}. We use $E_{\bfmu}^{\rm ef} \equiv \bigotimes_{m=1}^n \ketbra{\mu_m}{\mu_m}$ to denote the error-free POVM operator of the measurement outcome $\bfmu$. 

In our theoretical analysis, we assume that all single-qubit unitary gates are error-free. The overall map of the $j$-th-layer computing gates is $\calR_j \equiv [\bigotimes_{m=1}^n R_{(j-1)n+m}]$. Similarly, the overall map of the $j$-th-layer error-mitigating gates is $\calP_j \equiv [\bigotimes_{m=1}^n P_{(j-1)n+m}]$. Both $\calR_j$ and $\calP_j$ are error-free. Then, the error-free computing result is [see Fig.~\ref{fig:model}(a)] 
\begin{eqnarray}
&& {\rm com}^{\rm ef}(\bfR,\bfP) \notag \\
&=& \Tr\left[ E_f^{\rm ef} \left( \prod_{j=1}^{N+1} \calG_j^{\rm ef} \calP_{2j} \calR_j \calP_{2j-1} \right) \left(\rho_{\rm i}^{\rm ef}\right) \right],
\end{eqnarray}
where 
\begin{eqnarray}
E_f^{\rm ef} \equiv \sum_{\bfmu} f(\bfmu) E_{\bfmu}^{\rm ef},
\end{eqnarray}
$\calG_{N+1}^{\rm ef} = [\openone_{\rm S}]$ is the identity map, and $\openone_{\rm S}$ is the identity operator of $n$ qubits. Here, $\calR_j$ and $\calP_j$ depend on $\bfR$ and $\bfP$, respectively. 

In order to describe temporally-correlated errors, we introduce the environment in addition to the system (i.e.~$n$ qubits in the circuit). We use $\rho_{\rm i}$ to denote the initial state of the system and the environment. We use $\calG_j$ to denote the actual map acting on both the system and the environment for the $j$-th-layer frame gates. We use $E_{\bfmu}$ to denote the actual POVM operator of the system and the environment corresponding to the measurement outcome $\bfmu$. We define $\calR'_j \equiv \calR_j\otimes[\openone_{\rm E}]$ and $\calP'_j \equiv \calP_j\otimes[\openone_{\rm E}]$, where $\openone_{\rm E}$ is the identity operator of the environment. Then, the computing result with errors is [see Fig.~\ref{fig:model}(b)] 
\begin{eqnarray}
&& {\rm com}(\bfR,\bfP) \notag \\
&=& \Tr\left[ E_f \left( \prod_{j=1}^{N+1} \calG_j \calP'_{2j} \calR'_j \calP'_{2j-1} \right) \left(\rho_{\rm i}\right) \right],
\end{eqnarray}
where 
\begin{eqnarray}
E_f \equiv \sum_{\bfmu} f(\bfmu) E_{\bfmu},
\end{eqnarray}
and $\calG_{N+1} = [\openone_{\rm S}]\otimes[\openone_{\rm E}]$ is the identity map on both the system and the environment. 

\subsection{Tensor-product representation of quantum circuits}

\begin{figure}[tbp]
\centering
\includegraphics[width=1\linewidth]{\figpath 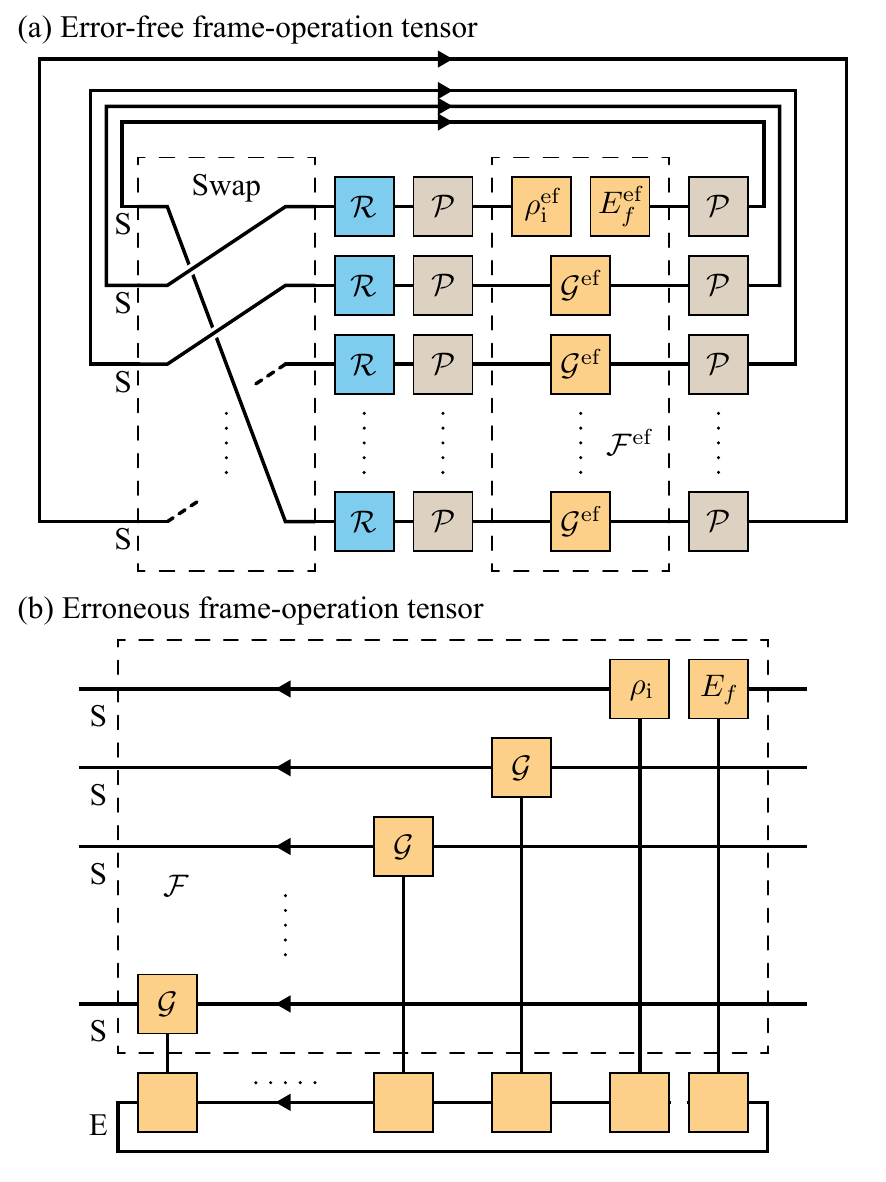}
\caption{
The error-free frame-operation tensor $\calF^{\rm ef}$ and the erroneous frame-operation tensor $\calF$. The arrows denote the direction of the time. Along the direction of arrows, $\calR = \calR_1,\calR_2,\ldots,\calR_{N+1}$, $\calP = \calP_1,\calP_2,\ldots,\calP_{2N+2}$, $\calG^{\rm ef} = \calG_1^{\rm ef},\calG_2^{\rm ef},\ldots\calG_N^{\rm ef}$ and $\calG = \calG_1,\calG_2,\ldots,\calG_{N}$. S: system; E: environment. 
}
\label{fig:tensors}
\end{figure}

Let $\{\ket{l}\}$ be the orthonormal basis of the Hilbert space. The trace of a map $\calM$ reads 
\begin{eqnarray}
\Tr\left( \calM \right) &\equiv & \sum_{l,l'} \Tr\left[ \ketbra{l'}{l} \calM \left(\ketbra{l}{l'}\right) \right].
\end{eqnarray}

We can express the identity map as 
\begin{eqnarray}
[\openone](\bullet) = \sum_{l,l'} \ketbra{l}{l'} \Tr\left( \ketbra{l'}{l} \bullet \right).
\end{eqnarray}
For arbitrary two maps $\calM_1$ and $\calM_2$, we have 
\begin{eqnarray}
&& \Tr\left(\calM_2\calM_1\right) = \Tr\left(\calM_2[\openone]\calM_1\right) \notag \\
&=& \sum_{l_1,l_1',l_2,l_2'} \Tr\left[\ketbra{l_1'}{l_1}\calM_2\left(\ketbra{l_2}{l_2'}\right)\right] \notag \\
&&\times \Tr\left[\ketbra{l_2'}{l_2}\calM_1(\ketbra{l_1}{l_1'})\right] \notag \\
&=& \sum_{l_1,l_1',l_2,l_2'} \Tr\left[ \ketbra{l_2'}{l_2}\otimes\ketbra{l_1'}{l_1} \right. \notag \\
&& \left. \calM_1\otimes\calM_2(\ketbra{l_1}{l_1'}\otimes\ketbra{l_2}{l_2'}) \right] \notag \\
&=& \sum_{l_1,l_1',l_2,l_2'} \Tr\left[ \ketbra{l_2'}{l_2}\otimes\ketbra{l_1'}{l_1} \calS_{1,2} \right. \notag \\
&& \left. \calM_2\otimes\calM_1(\ketbra{l_2}{l_2'}\otimes\ketbra{l_1}{l_1'}) \right] \notag \\
&=& \Tr\left( \calS_{1,2} \calM_2\otimes\calM_1 \right),
\end{eqnarray}
where $\calS_{1,2}$ is the swap map on two systems defined by 
\begin{eqnarray}
\calS_{1,2}(\bullet) &\equiv & \sum_{l_1,l_1',l_2,l_2'} \ketbra{l_1}{l_1'}\otimes\ketbra{l_2}{l_2'} \notag \\
&&\times \Tr\left(\ketbra{l_2'}{l_2}\otimes\ketbra{l_1'}{l_1}\bullet\right).
\end{eqnarray}
Similarly, for a product of $M$ maps, we have 
\begin{eqnarray}
&& \Tr\left( \calM_M\cdots\calM_2\calM_1 \right) \notag \\
&=& \Tr\left( \calS_{M-1,M} \calM_{M}\otimes\calM_{M-1}\cdots\calM_1 \right) \notag \\
&=& \Tr\left( \calS_{M-2,M-1}\calS_{M-1,M} \right. \notag \\
&& \left. \calM_M\otimes\calM_{M-1}\otimes\calM_{M-2}\cdots\calM_{1} \right) \notag \\
&=&\cdots \notag \\
&=& \Tr\left( \calS\calM_{M}\otimes\cdots\otimes\calM_2\otimes\calM_1 \right),
\end{eqnarray}
where $\calS \equiv \calS_{1,2}\calS_{2,3}\cdots\calS_{M-1,M}$. Here, we label the Hilbert spaces with $M,\ldots,2,1$ from left to right in the tensor product. 

\subsubsection{Error-free frame-operation tensor}

We introduce the map 
\begin{eqnarray}
\calG_0^{\rm ef}(\bullet) \equiv \rho_{\rm i}^{\rm ef}\Tr\left(E_f^{\rm ef} \bullet\right).
\end{eqnarray}
This map is linear, always Hermitian-preserving, trace-preserving if and only if $E_f^{\rm ef} = \openone$, completely positive if and only if $E_f^{\rm ef} \geq 0$. For an arbitrary map $\calM$ on the system, we have 
\begin{eqnarray}
\Tr\left( \calM \calG_0^{\rm ef} \right) &=& \sum_{\bfmu,\bfmu'} \Tr\left[ \ketbra{\bfmu}{\bfmu'} \calM \calG_0^{\rm ef} \left(\ketbra{\bfmu'}{\bfmu}\right) \right] \notag \\
&=& \Tr\left[ E_f^{\rm ef} \calM \left(\rho_{\rm i}^{\rm ef}\right) \right].
\end{eqnarray}

We define the error-free frame-operation tensor as 
\begin{eqnarray}
\calF^{\rm ef} \equiv \calG_N^{\rm ef}\otimes\cdots\otimes\calG_1^{\rm ef}\otimes\calG_0^{\rm ef},
\end{eqnarray}
which is a map on $N+1$ systems. Similarly, we define 
\begin{eqnarray}
\calR &\equiv & \calR_{N+1}\otimes\cdots\otimes\calR_2\otimes\calR_1, \notag \\
\calP_{\rm L} &\equiv & \calP_{2N+1}\otimes\cdots\otimes\calP_3\otimes\calP_1, \notag \\
\calP_{\rm R} &\equiv & \calP_{2N}\otimes\cdots\otimes\calP_2\otimes\calP_{2N+2}.
\end{eqnarray}
Let $\calS_{\rm S}$ be the swap map on $N+1$ systems. As shown in Fig.~\ref{fig:tensors}(a), we have 
\begin{eqnarray}
&& {\rm com}^{\rm ef}(\bfR,\bfP) \notag \\
&=& \Tr\left[ \left( \prod_{j=1}^{N+1} \calG_j^{\rm ef} \calP_{2j} \calR_j \calP_{2j-1} \right) \calG_0^{\rm ef} \right] \notag \\
&=& \Tr\left( \calS_{\rm S} \calS_{\rm S}^{-1}\calP_{\rm R}\calS_{\rm S} \calR \calP_{\rm L} \calF^{\rm ef} \right) \notag \\
&=& \Tr\left( \calS_{\rm S} \calR \calP_{\rm L} \calF^{\rm ef} \calP_{\rm R} \right).
\end{eqnarray}
Here, we have used that 
\begin{eqnarray}
\calS_{\rm S}^{-1}\calP_{\rm R}\calS_{\rm S} = \calP_{2N+2}\otimes\calP_{2N}\otimes\cdots\otimes\calP_2.
\end{eqnarray}

\subsubsection{Erroneous frame-operation tensor}

Similar to the error-free case, we define 
\begin{eqnarray}
\calG_0(\bullet) &=& \rho_{\rm i}\Tr\left(E_f \bullet\right), \notag \\
\calF' &\equiv & \calG_N\otimes\cdots\otimes\calG_1\otimes\calG_0, \notag \\
\calR' &\equiv & \calR'_{N+1}\otimes\cdots\otimes\calR'_2\otimes\calR'_1, \notag \\
\calP'_{\rm L} &\equiv & \calP'_{2N+1}\otimes\cdots\otimes\calP'_3\otimes\calP'_1, \notag \\
\calP'_{\rm R} &\equiv & \calP'_{2N}\otimes\cdots\otimes\calP'_2\otimes\calP'_{2N+2}.
\end{eqnarray}
Let $\calS_{\rm E}$ be the swap map on $N+1$ environments and $\calS' = \calS_{\rm S}\otimes\calS_{\rm E}$ be the swap map on $N+1$ system-environment composite systems.Then, 
\begin{eqnarray}
{\rm com}(\bfR,\bfP) = \Tr\left( \calS' \calR' \calP'_{\rm L} \calF' \calP'_{\rm R} \right).
\end{eqnarray}
Because $\calR' = \calR \otimes [\openone_{\rm E}]^{\otimes(N+1)}$, $\calP'_{\rm L} = \calP_{\rm L} \otimes [\openone_{\rm E}]^{\otimes(N+1)}$ and $\calP'_{\rm R} = \calP_{\rm R} \otimes [\openone_{\rm E}]^{\otimes(N+1)}$, we have 
\begin{eqnarray}
{\rm com}(\bfR,\bfP) = \Tr\left( \calS_{\rm S} \calR \calP_{\rm L} \calF \calP_{\rm R} \right),
\end{eqnarray}
where the erroneous frame-operation tensor, as shown in Fig.~\ref{fig:tensors}(b), is defined as 
\begin{eqnarray}
\calF \equiv \Tr_{\rm E} (\calS_{\rm E} \calF').
\end{eqnarray}

\section{Existence of a solution}
\label{app:existence}

Using the tensor-product representation, the error-mitigated computing result reads 
\begin{eqnarray}
{\rm com}^{\rm em}(\bfR,\bfI) = \Tr\left( \calS_{\rm S} \calR \calF^{\rm em} \right),
\end{eqnarray}
where the error-mitigated frame-operation tensor is 
\begin{eqnarray}
\calF^{\rm em} = \sum_{\bfP} q(\bfP) \calP_{\rm L} \calF \calP_{\rm R}.
\end{eqnarray}

It is straightforward to prove that the error-mitigated computing is error-free, i.e. 
\begin{eqnarray}
{\rm Error}(\bfR) = \absLR{ {\rm com}^{\rm em}(\bfR,\bfI) - {\rm com}^{\rm ef}(\bfR,\bfI) } = 0,
\end{eqnarray}
for all $\bfR\in\mathbb{U}$, if there exists a quasi-probability distribution $q(\bfP)$ satisfying 
\begin{eqnarray}
\sum_{\bfP} q(\bfP) \calP_{\rm L} \calF \calP_{\rm R} = \calF^{\rm ef}.
\end{eqnarray}
To solve this equation, we introduce the Pauli transfer matrix representation, i.e.~express maps using Pauli operators as the basis of the operator space. 

Let $\tau$ be the Pauli operator of $n(N+1)$ qubits. The Pauli transfer matrix of a $n(N+1)$-qubit map $\calM$ is 
\begin{eqnarray}
\overline{\calM}_{\tau_1,\tau_2} = 2^{-n(N+1)} \Tr\left[ \tau_1 \calM(\tau_2) \right].
\end{eqnarray}
Using Pauli transfer matrices, the equation becomes 
\begin{eqnarray}
\sum_{\bfP,\tau_2,\tau_3} q(\bfP) \overline{\calP}_{{\rm L};\tau_1,\tau_2} \overline{\calF}_{\tau_2,\tau_3} \overline{\calP}_{{\rm R};\tau_3,\tau_4} = \overline{\calF}^{\rm ef}_{\tau_1,\tau_4}.
\end{eqnarray}
The Pauli transfer matrix of a Pauli gate is always diagonal, i.e.~$\overline{\calP}_{{\rm L(R)};\tau_1,\tau_2} = \delta_{\tau_1,\tau_2} \overline{\calP}_{{\rm L(R)};\tau_1,\tau_1}$, where $\overline{\calP}_{{\rm L(R)};\tau_1,\tau_1} = \pm 1$. Therefore, we can rewrite the equation as 
\begin{eqnarray}
\sum_{\bfP} q(\bfP) \overline{\calP}_{{\rm L};\tau_1,\tau_1} \overline{\calF}_{\tau_1,\tau_4} \overline{\calP}_{{\rm R};\tau_4,\tau_4} = \overline{\calF}^{\rm ef}_{\tau_1,\tau_4}.
\end{eqnarray}
If $\overline{\calF}_{\tau_1,\tau_4}$ is nonzero for every nonzero element $\overline{\calF}^{\rm ef}_{\tau_1,\tau_4}$, we have 
\begin{eqnarray}
\sum_{\bfP} q(\bfP) \overline{\calP}_{{\rm L};\tau_1,\tau_1} \overline{\calP}_{{\rm R};\tau_4,\tau_4} = \overline{\calF}^{\rm ef}_{\tau_1,\tau_4}/\overline{\calF}_{\tau_1,\tau_4}.
\end{eqnarray}
For the $4^{2n(N+1)}$ error-mitigating gate sequences $\bfP$, the corresponding Pauli transfer matrices $\overline{\calP}_{\rm L} \otimes \overline{\calP}_{\rm R}$ are linearly-independent. Therefore, the solution of the equation always exists. 

One can check that Pauli transfer matrices of Pauli gates are linearly-independent diagonal matrices by computing Pauli transfer matrices of single-qubit Pauli gates. The Pauli transfer matrices of multi-qubit Pauli gates are tensor products of single-qubit matrices. 

\section{Information completeness}
\label{app:completeness}

We have proven the existence of a quasi-probability distribution $q(\bfP)$ satisfying 
\begin{eqnarray}
{\rm Loss} = \sum_{\bfR \in \mathbb{T}} {\rm Error}(\bfR)^2 = 0.
\end{eqnarray}
The training set $\mathbb{T}$ is information complete if ${\rm Error}(\bfR) = 0$ for all $\bfR\in \mathbb{U}$ when $q(\bfP)$ is a solution of ${\rm Loss} = 0$. 

The set $\mathbb{T} = \mathbb{C}$ containing all Clifford gate sequences is information complete. We only need to consider a subset of $\mathbb{C}$, which is $\mathbb{B} = \{ \bfR~\vert~R_j \in \mathbb{B}_1 \} \subset \mathbb{C}$, where $\mathbb{B}_1$ is a set of ten single-qubit Clifford gates 
\begin{eqnarray}
\mathbb{B}_1 &=& \{ I, X, Y, Z, \notag \\
&& (I+iX)/\sqrt{2}, (I+iY)/\sqrt{2}, (I+iZ)/\sqrt{2}, \notag \\
&& (Y+Z)/\sqrt{2}, (Z+X)/\sqrt{2}, (X+Y)/\sqrt{2} \}.
\end{eqnarray}
The maps of these ten Clifford gates are linearly independent. An arbitrary single-qubit unitary map $[R]$ can be decomposed as 
\begin{eqnarray}
[R] = \sum_{R'\in\mathbb{B}_1} \alpha_{R,R'} [R'].
\end{eqnarray}
Accordingly, 
\begin{eqnarray}
\calR = \sum_{\bfR'\in\mathbb{B}} \alpha_{\bfR,\bfR'} \calR',
\end{eqnarray} 
where $\alpha_{\bfR,\bfR'} = \prod_{j=1}^{n(N+1)} \alpha_{R_j,R'_j}$. 

When $\mathbb{T} = \mathbb{C}$, ${\rm Loss} = 0$ if and only if ${\rm Error}(\bfR) = 0$ for all $\bfR \in \mathbb{C}$, which means ${\rm com}^{\rm em}(\bfR,\bfI) = {\rm com}^{\rm ef}(\bfR,\bfI)$ for all $\bfR \in \mathbb{B}$. Then, we have 
\begin{eqnarray}
{\rm Error}(\bfR) &=& \absLR{ \sum_{\bfR'\in\mathbb{B}} \alpha_{\bfR,\bfR'} \left[ {\rm com}^{\rm em}(\bfR',\bfI) - {\rm com}^{\rm ef}(\bfR',\bfI) \right] } \notag \\
&=& 0
\end{eqnarray}
for all $\bfR\in \mathbb{U}$. Therefore, $\mathbb{T} = \mathbb{C}$ and $\mathbb{T} = \mathbb{B}$ are both information complete. 

\section{Fidelity measurement}
\label{app:fidelity}

The Pauli group of $n$ qubits is 
\begin{eqnarray}
P_n \equiv \{\pm 1,\pm i\}\times \{I,X,Y,Z\}^{\otimes n}.
\end{eqnarray}
The stabiliser group is a subgroup of the Pauli group, which reads 
\begin{eqnarray}
S \equiv \langle s_1,s_2,\cdots, s_n \rangle = \{ \prod_{i=1}^n s_i^{b_i} \},
\end{eqnarray}
where $s_i=s_i^\dag\in P_n$ are $n$ independent operators, $[s_i,s_j] = 0$ for all $i$ and $j$, and $b_i = 0,1$ are binary numbers. 

The stabiliser state $\ket{\psi_S}$ of the stabiliser group $S$ is the common eigenstate of all generators with the eigenvalue $+1$, i.e.~$s_i\ket{\psi_S} = \ket{\psi_S}$. The density matrix of the state can be written as 
\begin{eqnarray}
\rho_S = \ketbra{\psi_S}{\psi_S} = \prod_i \frac{\openone_{\rm S}+s_i}{2} = \frac{1}{2^n} \sum_{g\in S} g.
\end{eqnarray}
For a state $\rho$, the fidelity in the stabiliser state is 
\begin{eqnarray}
\bra{\psi_S} \rho \ket{\psi_S} = \Tr\left( \rho_S\rho \right) = \frac{1}{2^n} \sum_{g\in S} \Tr\left( g\rho \right).
\end{eqnarray}

\section{Pauli twirling and error model}
\label{app:Pauli}

We decompose error-mitigating gates into Pauli-twirling gates and error-correcting gates, i.e. 
\begin{eqnarray}
\calP_{\rm L} &=& \calP_{\rm L}^{\rm c} \calP_{\rm L}^{\rm t}, \\
\calP_{\rm R} &=& \calP_{\rm R}^{\rm t} \calP_{\rm R}^{\rm c},
\end{eqnarray}
where Pauli-twirling gates are 
\begin{eqnarray}
\calP_{\rm L}^{\rm t} &\equiv & \calP_G ^{\rm t} \otimes \calP_\rho ^{\rm t}, \\
\calP_{\rm R}^{\rm t} &\equiv & \left(\calG^{{\rm ef}-1}\calP_G ^{\rm t}\calG^{\rm ef}\right) \otimes \calP_E ^{\rm t},
\end{eqnarray}
and error-correcting gates are 
\begin{eqnarray}
\calP_{\rm L}^{\rm c} &\equiv & \calP_G ^{\rm c} \otimes \calP_\rho ^{\rm c}, \\
\calP_{\rm R}^{\rm c} &\equiv & [\openone_{\rm S}]^{\otimes N} \otimes \calP_E ^{\rm c}.
\end{eqnarray}
Here, the total frame gate is 
\begin{eqnarray}
\calG^{\rm ef} \equiv \calG_N^{\rm ef}\otimes\cdots\otimes\calG_1^{\rm ef}.
\end{eqnarray}

We define 
\begin{eqnarray}
\calP^{\rm t} &\equiv & \calP_E ^{\rm t} \otimes \calP_G ^{\rm t} \otimes \calP_\rho ^{\rm t}, \notag \\
\calP^{\rm c} &\equiv & \calP_E ^{\rm c} \otimes \calP_G ^{\rm c} \otimes \calP_\rho ^{\rm c},
\end{eqnarray}
which are $n(N+2)$-qubit Pauli gates. We use 
\begin{eqnarray}
\bfsigma &\equiv & \sigma_{2N+2}\otimes\sigma_{2N+1}\otimes\cdots\otimes\sigma_{3}\otimes\sigma_{1}
\end{eqnarray}
to denote a $n(N+2)$-qubit Pauli operator, where $\sigma_j$ are $n$-qubit Pauli operators. The Pauli-twirling gates are selected from the set 
\begin{eqnarray}
&& {\rm Twirling} \notag \\ &\equiv & \{I,Z\}^{\otimes n}\otimes\{I,X,Y,Z\}^{\otimes nN}\otimes\{I,Z\}^{\otimes n},
\end{eqnarray}
and error-correcting gates are selected from the set 
\begin{eqnarray}
&& {\rm Errors} \notag \\ &\equiv & \{I,X\}^{\otimes n}\otimes\{I,X,Y,Z\}^{\otimes nN}\otimes\{I,X\}^{\otimes n},
\end{eqnarray}
where $I,X,Y,Z$ are single-qubit Pauli operators. 
Then, $\calP^{\rm t} \in \{[\bfsigma] \st \bfsigma\in{\rm Twirling}\}$ and $\calP^{\rm c} \in \{[\bfsigma] \st \bfsigma\in{\rm Errors}\}$. 

To implement the Pauli twirling, we take $q(\bfP) = q_{\rm c}(\calP^{\rm c})/4^{n(N+1)}$, i.e.~$\calP^{\rm t}$ is uniformly distributed. Then, we have 
\begin{eqnarray}
&& \sum_{\bfP} q(\bfP) \calP_{\rm L} \calF \calP_{\rm R} \notag \\
&=& \sum_{\calP^{\rm c} \in \{[\bfsigma] \st \bfsigma\in{\rm Errors}\}} q_{\rm c}(\calP^{\rm c}) \calP_{\rm L}^{\rm c} \calF^{\rm pe} \calP_{\rm R}^{\rm c},
\end{eqnarray}
where the Pauli-error frame-operation tensor reads 
\begin{eqnarray}
\calF^{\rm pe} &=& \frac{1}{4^{n(N+1)}} \sum_{\calP^{\rm t} \in \{[\bfsigma] \st \bfsigma\in{\rm Twirling}\}} \calP_{\rm L}^{\rm t} \calF \calP_{\rm R}^{\rm t} \notag \\
&=& \sum_{\calP^{\rm e} \in \{[\bfsigma] \st \bfsigma\in{\rm Errors}\}} p(\calP^{\rm e}) \calP_{\rm L}^{\rm e} \calF^{\rm ef} \calP_{\rm R}^{\rm e}.
\label{eq:PE_frame}
\end{eqnarray}
Here, we have assumed that the measurement is balanced (see Sec.~\ref{app:BM}). See Sec.~\ref{app:PEM} for the proof. The Pauli errors are denoted by 
\begin{eqnarray}
\calP^{\rm e} &\equiv & \calP_E ^{\rm e} \otimes \calP_G ^{\rm e} \otimes \calP_\rho ^{\rm e}, \\
\calP_{\rm L}^{\rm e} &\equiv & \calP_G ^{\rm e} \otimes \calP_\rho ^{\rm e}, \\
\calP_{\rm R}^{\rm e} &\equiv & [\openone_{\rm S}]^{\otimes N} \otimes \calP_E ^{\rm e},
\end{eqnarray}
$p(\calP^{\rm e})\geq 0$ is the probability of the error, and $\sum_{\calP^{\rm e} \in \{[\bfsigma] \st \bfsigma\in{\rm Errors}\}} p(\calP^{\rm e}) = 1$. 

\subsection{Balanced measurement}
\label{app:BM}

We use $X_{\bfb} = \bigotimes_{m=1}^n X^{b_m}$ to denote an $n$-qubit Pauli operator, where $\bfb = (b_1,b_2,\ldots,b_n)$ is a binary vector. The balanced measurement is defined as a measurement that satisfies $E_{\bfmu\oplus\bfb} = [X_{\bfb}\otimes\openone_{\rm E}]\left(E_{\bfmu}\right)$, where $\bfmu\oplus\bfb = (\mu_1+b_1,\mu_2+b_2,\ldots,\mu_n+b_n)~{\rm mod}~2$. 

For a balanced measurement, we have $E_{\bfmu} = [X_{\bfmu}\otimes\openone_{\rm E}]\left(E_{\bfzero}\right)$, where $\bfzero = (0,0,\ldots,0)$. Because $\sum_{\bfmu} E_{\bfmu} = \sum_{\bfmu} [X_{\bfmu}\otimes\openone_{\rm E}]\left(E_{\bfzero}\right) = \openone_{\rm S}\otimes\openone_{\rm E}$, $E_{\bfzero}$ satisfies $\Tr_{\rm S}(E_{\bfzero}) = \openone_{\rm E}$. 

Under the condition that Pauli gates are error-free, an arbitrary raw measurement with POVM operators $\{ E_{\bfmu}^{\rm raw} \}$ can be converted into a balanced measurement by randomly applying the gate $X_{\bfb}$ before the measurement and record the outcome taking into account the applied gate, i.e.~record the outcome as $\bfmu$ if the raw measurement outcome is $\bfmu\oplus\bfb$. As a result, POVM operators of the effective measurement is $E_{\bfmu} = 2^{-n}\sum_{\bfb} [X_{\bfb}\otimes\openone_{\rm E}] \left(E_{\bfmu\oplus\bfb}^{\rm raw}\right)$. One can find that $\{ E_{\bfmu} \}$ is a balanced measurement. 

\subsection{Pauli error model}
\label{app:PEM}

In this section we prove Eq.~(\ref{eq:PE_frame}). 

Let $\rho_{\rm i} = \sum_{\bfa,\bfb} \ketbra{\bfa}{\bfb}\otimes\rho_{{\rm E};\bfa,\bfb}$ be the initial state, where $\bfa = (a_1,a_2,\ldots,a_n)$ is a binary vector, and $\ket{\bfb} = X_{\bfb}\ket{0}^{\otimes n} = \bigotimes_{m=1}^n \ket{b_m}$. Here, $\rho_{{\rm E};\bfa,\bfb}$ are matrices acting on the Hilbert space space of the environment and satisfy $\rho_{{\rm E};\bfa,\bfb}^\dag = \rho_{{\rm E};\bfb,\bfa}$, $\rho_{{\rm E};\bfb,\bfb}\geq 0$ and $\Tr(\rho_{\rm E}) = 1$, where the initial state of the environment $\rho_{\rm E} = \sum_{\bfb}\rho_{{\rm E};\bfb,\bfb}$. By applying the twirling gates, we get the effective initial state 
\begin{eqnarray}
\rho_{\rm eff} &=& \left(\frac{[I]+[Z]}{2}\right)^{\otimes n}\otimes[\openone_{\rm E}] \left(\rho_{\rm i}\right) = \sum_{\bfb} \ketbra{\bfb}{\bfb}\otimes\rho_{{\rm E};\bfb,\bfb} \notag \\
&=& \sum_{\bfb} [X_{\bfb}] \left(\rho_{\rm i}^{\rm ef}\right) \otimes\rho_{{\rm E};\bfb,\bfb},
\end{eqnarray}
where $\rho_{\rm i}^{\rm ef} = \ketbra{\bfzero}{\bfzero}$. 

For a balanced measurement $\{ E_{\bfmu} \}$, we have $E_f = \sum_{\bfmu} f(\bfmu) E_{\bfmu} = \sum_{\bfmu} f(\bfmu) [X_{\bfmu}\otimes\openone_{\rm E}] \left(E_{\bfzero}\right)$. Similar to the state, we can express the POVM operator as $E_{\bfzero} = \sum_{\bfa,\bfb} \ketbra{\bfa}{\bfb}\otimes E_{{\rm E};\bfa,\bfb}$.  Here, $E_{{\rm E};\bfa,\bfb}$ are matrices acting on the Hilbert space space of the environment and satisfy $E_{{\rm E};\bfa,\bfb}^\dag = E_{{\rm E};\bfb,\bfa}$, $E_{{\rm E};\bfb,\bfb}\geq 0$ and $E_{\rm E} = \sum_{\bfb} E_{{\rm E};\bfb,\bfb} = \openone_{\rm E}$. By applying the twirling gates, we get the effective POVM operator 
\begin{eqnarray}
E_{\rm eff} &=& \left(\frac{[I]+[Z]}{2}\right)^{\otimes n}\otimes[\openone_{\rm E}] \left(E_f\right) \notag \\
&=& \sum_{\bfmu} f(\bfmu) [X_{\bfmu}] \left(\sum_{\bfb} \ketbra{\bfb}{\bfb}\right)\otimes E_{{\rm E};\bfb,\bfb} \notag \\
&=& \sum_{\bfb} [X_{\bfb}] \left(\sum_{\bfmu} f(\bfmu) \ketbra{\bfmu}{\bfmu}\right)\otimes E_{{\rm E};\bfb,\bfb} \notag \\
&=& \sum_{\bfb} [X_{\bfb}] \left(E_f^{\rm ef}\right)\otimes E_{{\rm E};\bfb,\bfb}.
\end{eqnarray}

For a gate $\calG_j$, because $\calG_j^{\rm ef}$ is a unitary map, we can always rewrite it in the form $\calG_j = \calN_j (\calG_j^{\rm ef}\otimes[\openone_{\rm E}])$, where $\calN_j$ is the noise map acting on both the system and the environment, which is completely positive and trace-preserving. By applying the twirling gates, we get the effective gate 
\begin{eqnarray}
&& \calG_{{\rm eff};j} \notag \\
&=& \frac{1}{4^n} \sum_{\sigma\in \{I,X,Y,Z\}^{\otimes n}} \left([\sigma]\otimes[\openone_{\rm E}]\right) \calG_j \left(\calG_j^{{\rm ef}-1}[\sigma]\calG_j^{\rm ef}\otimes[\openone_{\rm E}]\right) \notag \\
&=& \calN_{{\rm eff};j} (\calG_j^{\rm ef}\otimes[\openone_{\rm E}]),
\end{eqnarray}
where the effective noise map 
\begin{eqnarray}
\calN_{{\rm eff};j} &=& \frac{1}{4^n} \sum_{\sigma\in \{I,X,Y,Z\}^{\otimes n}} \left([\sigma]\otimes[\openone_{\rm E}]\right) \calN_j \left([\sigma]\otimes[\openone_{\rm E}]\right) \notag \\
&=& \sum_{\sigma} [\sigma]\otimes\calN_{{\rm E};j,\sigma},
\label{eq:noise_eff}
\end{eqnarray}
$\calN_{{\rm E};j,\sigma}$ are completely positive maps acting on the environment, and $\calN_{{\rm E};j} = \sum_{\sigma} \calN_{{\rm E};j,\sigma}$ is trace-preserving. 

To prove Eq.~(\ref{eq:noise_eff}), we consider a completely positive map acting on one qubit and an ancillary system A. The map reads $\calM = \sum_K [K]$, and $K = \sum_{P=I,X,Y,Z} P\otimes K_P$, where $\{K_P\}$ are matrices acting on the ancillary system. The effective map with the Pauli twirling reads 
\begin{eqnarray}
\calM_{\rm eff} &=& \frac{1}{4} \sum_{P=I,X,Y,Z} \left([P]\otimes[\openone_{\rm A}]\right) \calM \left([P]\otimes[\openone_{\rm A}]\right) \notag \\
&=& \sum_K \sum_P [P]\otimes[K_P] = \sum_P [P]\otimes\calM_P,
\end{eqnarray}
where $\openone_{\rm A}$ is the identity operator of the ancillary system, and $\calM_P = \sum_K [K_P]$ is a completely positive map acting on the ancillary system. Because $\sum_K K^\dag K = \sum_K\sum_{P,P'} PP'\otimes K_P^\dag K_{P'}$, we have $\sum_K\sum_P K_P^\dag K_P = \frac{1}{2}\Tr_{\rm qubit}\left(\sum_K K^\dag K\right)$. Therefore, if $\calM$ is trace-preserving, $\sum_P \calM_P$ is also trace-preserving. By applying this approach to qubits one by one, we can obtain Eq.~(\ref{eq:noise_eff}). 

Now, we can see that the frame-operation tensor with the Pauli twirling is in the Pauli-error form, as given in Eq.~(\ref{eq:PE_frame}). For the Pauli error $\calP^{\rm e} = [\bfsigma] = [X_{\bfb}\otimes\sigma_{2N+1}\otimes\cdots\otimes\sigma_{3}\otimes X_{\bfa}]$, the corresponding error probability is 
\begin{eqnarray}
&& p([\bfsigma]) \notag \\
&=& \Tr\left( E_{{\rm E};\bfb,\bfb}\calN_{{\rm E};N,\sigma_{2N-1}}\cdots\calN_{{\rm E};2,\sigma_3}\calN_{{\rm E};1,\sigma_1}\rho_{{\rm E};\bfa,\bfa} \right).~~~
\end{eqnarray}
We have $p([\bfsigma])\geq 0$, because $\rho_{{\rm E};\bfa,\bfa}$ and $E_{{\rm E};\bfb,\bfb}$ are positive, and $\calN_{{\rm E};j,\sigma}$ are completely positive. We also have 
\begin{eqnarray}
&& \sum_{[\bfsigma] \st \bfsigma\in{\rm Errors}} p([\bfsigma]) \notag \\
&=& \Tr\left( E_{\rm E}\calN_{{\rm E};N}\cdots\calN_{{\rm E};2}\calN_{{\rm E};1}\rho_{\rm E} \right) = 1,~~~
\end{eqnarray}
because $\rho_{\rm E}$ is normalised, $E_{\rm E}$ is identity, and $\calN_{{\rm E};j}$ are trace-preserving. 

\section{Example of SigE generation}
\label{app:Example}

Let us start with the set of all possible circuit variations $S = \{\bfsigma\} $ described by the pattern of Pauli errors $\bfsigma = \sigma_1\otimes\sigma_3\otimes\cdots\otimes\sigma_{2N+1}\otimes\sigma_{2N+2} = (\sigma_1, \sigma_2, ..., \sigma_{2N+2})$, i.e. $S = \{(I^{\otimes n}, I^{\otimes n}, ..., I^{\otimes n}), (I^{\otimes n}, I^{\otimes n}, ..., I^{\otimes n-1}\otimes X), ...,  (X^{\otimes n}, Z^{\otimes n}, ..., X^{\otimes n}), ...\}$.
Assume the gate set tomography perfectly identifies or we have pre-existing knowledge of the local two-qubit depolarising noise after each application of a two-qubit gate, 
\begin{equation}
\calD_{\rm Pol} = (1-\epsilon)[\openone] + \frac{\epsilon}{15}\sum_{\mu\in \{I,X,Y,Z\}^{\otimes 2}\setminus I^{\otimes 2}} [\mu],
\end{equation}
with some severity $\epsilon \in [0, \frac{15}{16}]$. Inverting this map yields
\begin{equation}
\calD_{\rm Pol}^{-1} = \eta_1[\openone] + \eta_2 \sum_{\mu\in \{I,X,Y,Z\}^{\otimes 2}\setminus I^{\otimes 2}} [\mu],
\end{equation}
where $\eta_1 = 1 + 15\epsilon(15-16\epsilon)^{-1}$, $\eta_2 = -\epsilon(15-16\epsilon)^{-1}$ and $|\eta_1|>|\eta_2|$ (Note that for $\epsilon = \frac{15}{16}$ the map is not invertible). In the error mitigated computation according to this error model, after each two-qubit gate in the circuit, we either apply $[\openone]$ with probability $|\eta_1|/\gamma$ or each $[\mu] \in \{I,X,Y,Z\}^{\otimes 2}\setminus I^{\otimes 2}$ with probability $|\eta_2|/\gamma$ in each run of the circuit, with $\gamma = |\eta_1| + 15|\eta_2|$ being the overhead factor. 

Now consider that we have $P$ noisy two-qubit gates in the circuit. In the error-mitigated computation we can also sample circuits according to their quasi-probability distribution. We have $|\eta_1^P|/\gamma^P$ chance to run a circuit variation $\bfsigma = (I^{\otimes n}, I^{\otimes n}, ..., I^{\otimes n})$, $|\eta_1^{P-1}\eta_2|/\gamma^P$ chance to run a circuit with some $[\mu]$ applied after one of the two-qubit gates, but nowhere else, e.g $\bfsigma = (I^{\otimes n}, I^{\otimes n}, ..., I^{\otimes n-1}\otimes X, I^{\otimes n})$; $|\eta_1^{P-2}\eta_2^2|/\gamma^P$ chance to apply some $[\mu]$ only after two two-qubit gates, but nowhere else, etc. 
In this example, these are the variations of the circuit with non-zero initial quasi-probability $q(\bfsigma)^{\mathrm{ini}} \neq 0$ and they don't necessarily form the full set $S$. The first step in the SigE construction filters out all other variations for which $q(\bfsigma)^{\mathrm{ini}} = 0$. For example, $\bfsigma = (I^{\otimes n-1}\otimes X, I^{\otimes n}, ..., I^{\otimes n}, I^{\otimes n})$, because there is a non-identity gate applied after assumed perfect initialisation.

The second step truncates the set SigE by excluding variations with the lowest chance of being selected when randomly picking one of the circuit variations. For example, circuit with $\sigma_{2j+1}= Z^{\otimes n}$ $\forall j$ has an order constant $k=P$, there are non-identity Pauli gate/s directly after all $P$ two-qubit gates, and all circuits with that order constant have a probability $|\eta_2^P|/\gamma^P$ of being randomly chosen. In this step we exclude all circuit variations with lowest probabilities up to some order constant $k=z$, meaning that all variations in ${\rm SigE}$ will have at least probability $|\eta_1^{P-z}\eta_2^z|/\gamma^P$ of being implemented. In our simulations we have set $k=1$.

In this way we neglect the lowest chance variations of the circuit in the optimisation stage of the protocol, and by doing that we limit the number of quasi-probabilities that we need to optimise to a polynomially scaling number with the circuit size.

\section{Circuit layout}
\label{app:circuitL}

\begin{figure}[tbp]
\centering
\includegraphics[width=1\linewidth]{\figpath 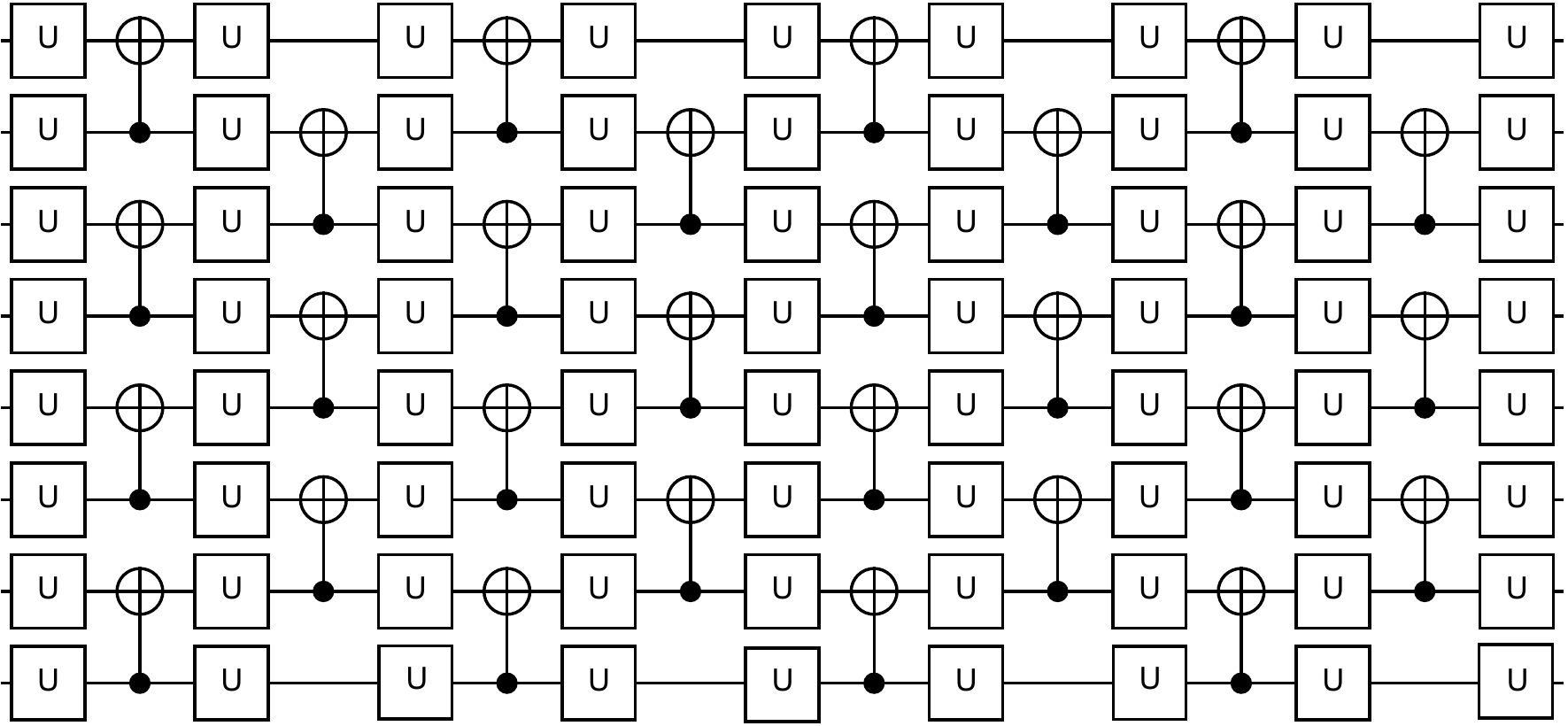}
\caption{
Eight-qubit wide and eight-layer deep circuit layout. Gates U represent single qubit unitary gates, and the two-qubit gates are controlled-NOT gates. For circuits in $\mathbb{T}$, gates U are all Clifford. The initial state is $\ket{0}^{\otimes n}$, and the measurement is done in the $Z$ basis on the bottom qubit. 
}
\label{fig:circuitL}
\end{figure}

See Fig.~\ref{fig:circuitL} for the circuit layout used in our simulations. 

\section{Error model for numerical simulations}
\label{app:EM}

All noisy quantum circuits share the same base error model - every controlled-NOT gate with control qubit $i$ and target qubit $i+1$ is followed by a two-qubit channel $\calD$ acting on qubits $i$ and $i+1$. Here $\calD$ represents either two-qubit depolarising channel 
\begin{equation}
\calD_{\rm Pol} = (1-\epsilon)[\openone] + \frac{\epsilon}{15}\sum_{\mu\in \{I,X,Y,Z\}^{\otimes 2}\setminus I^{\otimes 2}} [\mu]
\end{equation}
or two-qubit dephasing channel 
\begin{equation}
\calD_{\rm Ph} = (1-\epsilon)[\openone] + \frac{\epsilon}{3}\sum_{\mu\in \{I,Z\}^{\otimes 2}\setminus I^{\otimes 2}} [\mu],
\end{equation}
with the error rate $\epsilon = 0.01$. To incorporate spatially or temporally correlated errors, which are partially/fully unnoticed in the two-qubit tomography, we modify the base error model in two ways separately: 

\textbf{A}. After each channel $\calD_{\rm x}$ on qubits $i$ and $i+1$, we apply another \textit{two} channels $\calD_{\rm x}$ with the same error rate $\epsilon$ on qubits $i+1$ and $i+2 \mod n$ and qubits $i-1 \mod n$ and $i$. Here $x$ may denote depolarising $x={\rm Pol}$ or dephasing channel $x={\rm Ph}$. Note periodic boundary conditions, i.e.~qubit 1 can cross-talk to qubit $n$. Here, the sequencing for a two-qubit gate layer is such that after each ideal gate, the three $\calD_{\rm x}$ noise channels are implemented before the next two-qubit ideal gate. Gates in the same layer are implemented from the top one to the bottom one, Fig.~\ref{fig:circuitL}.

\textbf{B}. Every time the circuit is run, a single qubit $i$, following a probability distribution ${\rm Prob}(i)$, has a chance to be worse than other qubits. Meaning that every channel $\calD_x$, the qubit $i$ is part of, has an increased error rate $\epsilon^* = g\epsilon$. In our numerical simulations we use a uniform distribution ${\rm Prob}(i)$ and set $g=10$.

\section{Sampling of Clifford circuits}
\label{app:Clif}

We consider the case that the circuit is for measuring the mean value of a physical quantity $E^{\rm ef}_f$, which is a Pauli operator. 

For a Clifford circuit, the final state is a stabiliser state, i.e.~the eigenstate of a set of Pauli operators. These Pauli operators generates the stabiliser group. If $E^{\rm ef}_f$ commutes with all stabiliser operators, it is an element of the stabiliser group up to a sign. In this case, the mean value of $E^{\rm ef}_f$ is either $+1$ or $-1$. If $E^{\rm ef}_f$ anti-commutes with any stabiliser operator, the mean value of $E^{\rm ef}_f$ is $0$. Pauli errors in the circuit do not change the stabiliser group but flip eigenvalues. Therefore, given a Pauli error configuration, the mean value of $E^{\rm ef}_f$ may be flipped from $+1$ to $-1$ or from $-1$ to $+1$. If the error-free mean value is $0$, Pauli errors do not change it. 

We consider the example with only one qubit and one gate. The qubit is initialised in the state $\ket{0}$, a Clifford gate $R$ is performed on the qubit, and we measure $E^{\rm ef}_f = Z$. If $R = I$, $\mean{Z} = 1$; if $R = H$, $\mean{Z} = 0$; and if $R = X$, $\mean{Z} = -1$. If there is an $X$ error on the qubit, which occurs just before the measurement, then we have: If $R = I$, $\mean{Z} = -1$; if $R = H$, $\mean{Z} = 0$; and if $R = X$, $\mean{Z} = 1$. If the $X$ error occurs with the probability $p$, we have: If $R = I$, $\mean{Z} = 1-2p$; if $R = H$, $\mean{Z} = 0$; and if $R = X$, $\mean{Z} = -(1-2p)$. 

We can find, given a Clifford circuit $\bfR$, if ${\rm com}^{\rm ef}(\bfR,\bfI) = 0$, we always have ${\rm com}(\bfR,\bfP) = 0$ and ${\rm com}^{\rm em}(\bfR,\bfI) = 0$. Therefore, such a circuit does not contribute to the loss function. According to the importance sampling, when we compute the loss function, we only need to sample Clifford sequences $\bfR$ with ${\rm com}^{\rm ef}(\bfR,\bfI) = \pm 1$. In the numerical simulations, Clifford circuits are randomly generated and selected in this way. 

\section{Clifford circuit overhead}
\label{app:cScale}
\begin{figure}[tbp]
\centering
\includegraphics[width=1\linewidth]{\figpath 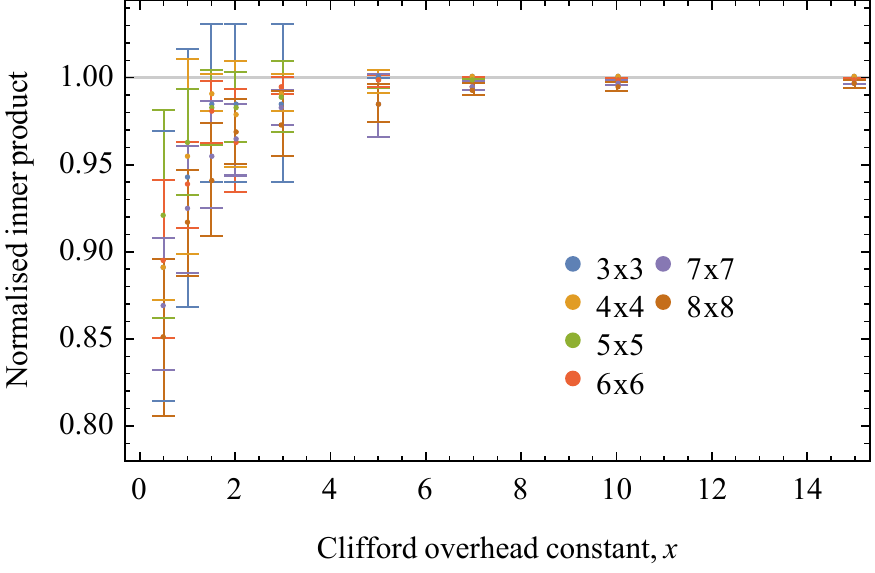}
\caption{
Normalised inner product $q_x.q_{20}/(|q_x||q_{20}|)$ between quasi-probability distributions that are obtained in the learning part of the protocol, as described in Section~\ref{sec:SigE}, with Clifford overhead constant $c=x$ ($q_x$) and $c=20$ ($q_{20}$). We plot the inner product for different circuit sizes $n{\rm x}n$ where $n$ is both the number of qubits and the number of layers of two-qubit gates. Here we have chosen $q_{20}$ as the reference for other distributions assuming that $q_{20}$ deviates only marginally from $q'_{\rm opt}$ which is obtained from the full training set $\mathbb{T}$. This allows us to estimate the impact on errors due to the truncation of the training set and, hence, let us choose a suitable Clifford overhead constant $c$ for our simulations.
}
\label{fig:cliffS}
\end{figure}

Here we present a numerical study showing the effect of different Clifford overhead constants $c$ for circuits up to 8 qubits using the significant-error approach(Fig.~\ref{fig:cliffS}). While no asymptotic scaling for $c$ can be undeniably determined with respect to the circuit size, our data suggest that for small systems a Clifford overhead constant $c=7$ is sufficient to introduce only negligible errors due to the truncation of the training set $\mathbb{T}$. In our simulations we chose $c=3$, since the error due to a finite sampling (shot noise) dominates the error due to the size of the training set being $|\mathbb{T}|=3|{\rm SigE}|$.

\section{Single-parameter optimisation}
\label{app:SinglePO}

For single-parameter learning  we optimise $q_{\rm }(\bfsigma)_{\rm ini}$ $\forall \bfsigma \in {\rm SigE}$ generated with $k=1$ where the optimisation is constrained to a single adjustable parameter.
The severity of the local noise $\epsilon$ is chosen as the parameter and, hence, the respective $q(\bfsigma)$ is then classically derived by inverting the local noise channel (see Appendix~\ref{app:Example}). Finding $q_{\rm opt}(\bfsigma)$ is equivalent to finding $\epsilon_{\rm opt}$. The lower bound on performance of single-parameter learning is then set by tomography-based error mitigation with $q_{\rm }(\bfsigma)_{\rm ini}$ generated with $k=1$ assuming the optimiser can always find the global minima.

As an example, here we present results for a single-parameter learning compared to a multi-parameter one for a 7 qubit and 7 depth circuit with spatially correlated dephasing errors described in Appendices \ref{app:circuitL} and \ref{app:EM}, Fig. \ref{fig:Single-para}.

\begin{figure}[tbp]
\centering
\includegraphics[width=1\linewidth]{\figpath 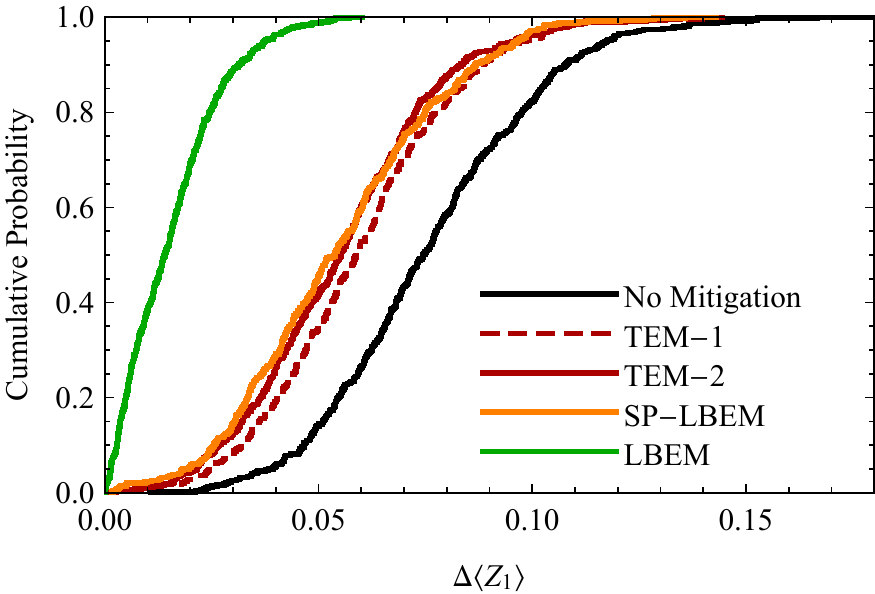}
\caption{
Empirical cumulative distribution function of estimated $\Delta\mean{Z_1}$ for 500 pseudo-random circuits with spatially correlated dephasing noise. Results for circuits without error mitigation (black), with tomographic error mitigation with $k=1$ (dashed red) and $k=2$ (red), with a single-parameter learning based error mitigation (orange) and with a multi-parameter learning-based error mitigation (green) are presented.
}
\label{fig:Single-para}
\end{figure}

The numerical results indicate that single-parameter learning just marginally outperforms its lower bound and is comparable to a tomography-based error mitigation with $q_{\rm }(\bfsigma)_{\rm ini}$ generated with $k=2$. Results for the other two error models follow suit.

\section{Hardware efficient variational circuit}
\label{app:circuitLNew}

\begin{figure*}[tbp]
\centering
\includegraphics[width=1\linewidth]{\figpath 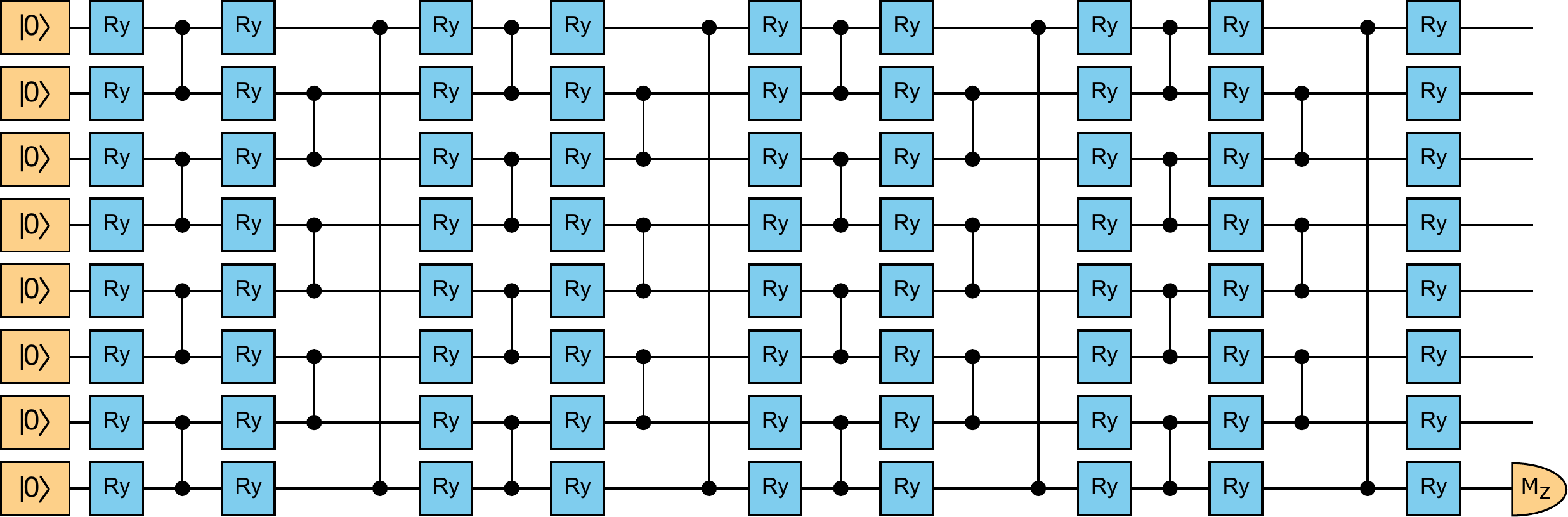}
\caption{
Qubits follow the cycle graph pattern with the first qubit being adjacent to the last one. Gates are implemented from bottom up if they are on the same vertical line. Black two-qubit gates are control-Z gates and single qubit $R_y$ gates are rotations around $y$ axis of the bloch sphere.
}
\label{fig:circuitLNew}
\end{figure*}

See Fig.~\ref{fig:circuitLNew} for hardware efficient variational circuit used in our numerical study.

\section{Evaluating and minimising the loss function for the product-form ansatz}
\label{app:LossFn}

The loss function reads 
\begin{eqnarray}
{\rm Loss} &=& \frac{1}{\abs{\mathbb{T}}} \sum_{\bfR\in\mathbb{T}}\vert {\rm com}^{\rm em}(\bfR,\bfI)-{\rm com}^{\rm ef}(\bfR,\bfI)\vert^{2}.
\end{eqnarray}
We define two functions, the quasi-probability function 
\begin{eqnarray}
V(\bfq,\bfb) &=& \prod_{i=1}^{\abs{\rm SigE}} \left[ b_iq_i + (1-b_i)(1-q_i) \right],
\end{eqnarray}
and the probability function 
\begin{eqnarray}
W(\bfq,\bfb) &=& \prod_{i=1}^{\abs{\rm SigE}} \frac{b_i\abs{q_i} + (1-b_i)\abs{1-q_i}}{\abs{q_i} + \abs{1-q_i}}.
\end{eqnarray}

\begin{table}[tbp]
\begin{tabular}{|c|c|c|c|}
\hline 
Circuit size & $\gamma^{\prime}$ & Circuit size & $\gamma^{\prime}$\tabularnewline
\hline 
\hline 
$5\times5$ & $10^{-4}$ & $11\times11$ & $7\times10^{-5}$\tabularnewline
\hline 
$6\times6$ & $10^{-4}$ & $12\times12$ & $7\times10^{-5}$\tabularnewline
\hline 
$7\times7$ & $9\times10^{-5}$ & $15\times15$ & $5\times10^{-5}$\tabularnewline
\hline 
$8\times8$ & $9\times10^{-5}$ & $16\times16$ & $5\times10^{-5}$\tabularnewline
\hline 
$9\times9$ & $8\times10^{-5}$ & $19\times19$ & $3\times10^{-5}$\tabularnewline
\hline 
$10\times10$ & $8\times10^{-5}$ & $20\times20$ & $3\times10^{-5}$\tabularnewline
\hline 
\end{tabular}
\caption{
Learning rate used for different size circuits. Circuit sizes are described in a short-hand notation by $n \times N$ with $n$ qubits and $N$ layers.
}
\label{tableII}
\end{table}

Given a configuration of computing gates $\bfR$, to evaluate the error-mitigated result ${\rm com}^{\rm em}(\bfR,\bfI)$, we randomly generate $M$ configurations of error-correcting gates $\bfP$. We can label each of them with $\bfb_k$, where $k=1,2,\ldots,M$, and the corresponding configuration of error-correcting gates is $\bfP_{\bfb_k}$. These gate configurations are randomly generated according to the distribution $W(\bfp,\bfb)$. We always take $\bfp = \bfq$ in order to minimise the variance. Then, we compute 
\begin{eqnarray}
\widehat{{\rm com}}^{\rm em}(\bfR,\bfI) = \frac{1}{M} \sum_{k=1}^M \frac{V(\bfq,\bfb_k)}{W(\bfp,\bfb_k)} f_k, 
\end{eqnarray}
where $f_k$ is the value of the observable obtained in one shot of the circuit $(\bfR,\bfP_{\bfb_k})$. To evaluate the loss function ${\rm Loss}(\bfq)$, we randomly generate $N$ configurations of computing gates $\bfR$ with non-zero error-free computing result, i.e.~${\rm com}^{\rm ef}(\bfR,\bfI) = \pm 1$. We label these $N$ configurations with $\bfR_j$, where $j=1,2,\ldots,N$. Given the error mitigated result of each $\bfR_j$, we compute 
\begin{eqnarray}
\widehat{{\rm Loss}} &=& \frac{1}{N} \sum_{j=1}^N \vert\widehat{{\rm com}}^{\rm em}(\bfR_j,\bfI)-{\rm com}^{\rm ef}(\bfR_j,\bfI)\vert^{2}.
\end{eqnarray}
$\widehat{{\rm com}}^{\rm em}(\bfR,\bfI)$ and $\widehat{{\rm Loss}}$ are estimators of ${\rm com}^{\rm em}(\bfR,\bfI)$ and ${\rm Loss}$, respectively. 

\begin{figure}[tbp]
\begin{center}
\includegraphics[width=1\linewidth]{\figpath/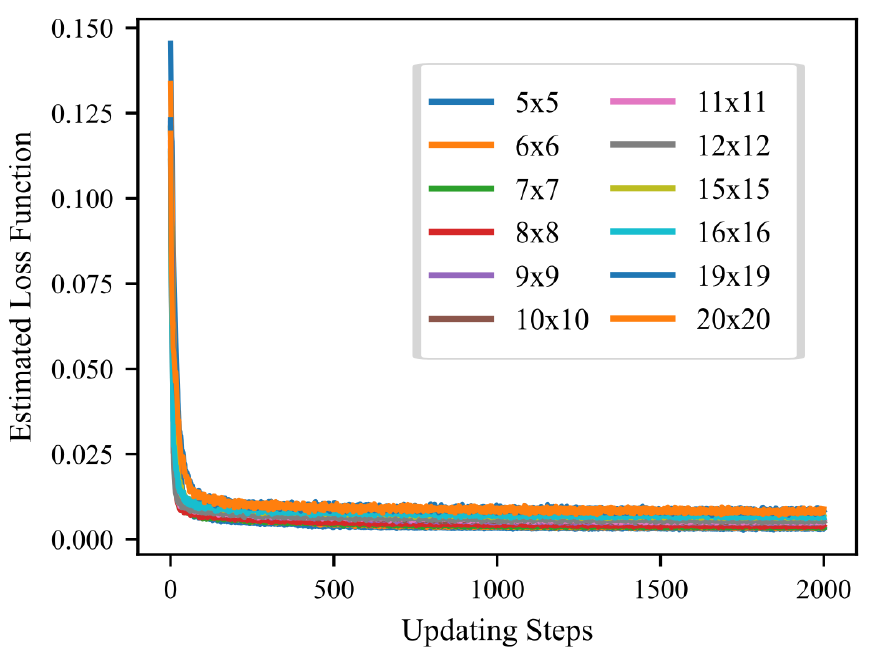}
\caption{
Values of the loss function in the gradient descent. We label an $n$-qubit $N$-layer circuit with $n\times N$. For each circuit size, we randomly generate $N = 1000$ computing-gate configurations $\bfR$, and for each computing-gate configuration, we randomly generate $M=1000$ configurations of error-correcting gates $\bfP$. Note that the estimated cost function does not converge to zero with a finite number of samples due to a biased estimator. 
}
\label{fig:loss_pf}
\end{center}
\end{figure}

To minimise the loss function, we compute the gradient of the loss function with respect to the quasi-probability $\bfq$, i.e. 
\begin{eqnarray}
\frac{\partial \widehat{{\rm Loss}}}{\partial q_i} &=& \frac{2}{N} \sum_{j=1}^N \left[\widehat{{\rm com}}^{\rm em}(\bfR_j,\bfI)-{\rm com}^{\rm ef}(\bfR_j,\bfI)\right] \notag \\
&&\times \frac{1}{M} \sum_{k=1}^M \frac{\partial V(\bfq,\bfb_k)/\partial q_i}{W(\bfp,\bfb_k)} {\rm com}(\bfR_j,\bfP_{\bfb_k}).
\end{eqnarray}
Then, we update the quasi-probabilities according to 
\begin{equation}
q_i \leftarrow q_i - \gamma \frac{\partial \widehat{{\rm Loss}}}{\partial q_i},
\end{equation}
where $\gamma$ is the learning rate. To make sure that parameters are updated at a reasonable level, a trick we used for gradient descent is using a dynamical learning rate 
\begin{equation}
\gamma = \frac{\max\{\abs{q_i-1}\}}{\max\{ \abs{\partial \widehat{{\rm Loss}}/\partial q_i} \}}\gamma',
\end{equation}
and we take a fixed value of $\gamma^{\prime}$, which is listed in Table.~\ref{tableII}. The decreasing of estimated loss functions are plotted in Fig.~\ref{fig:loss_pf}. 

\section{Monte Carlo summation}
\label{app:MC}

We consider two cases. In the first case, the quasi-probability $q(\bfP)$ is non-zero only if $\bfP \in {\rm SigE}$, where ${\rm SigE}$ is the set of significant Pauli errors including the trivial error $\bfI$, and the value of each $q(\bfP)$ is the variational parameter, i.e.~the number of parameters is $\abs{\rm SigE}$. In the second case, the quasi-probability is expressed as 
\begin{eqnarray}
q(\bfP) = C \frac{B(\bfP,\lambda)}{A(\lambda)},
\end{eqnarray}
where $B(\bfP,\lambda)$ is a real-valued function with an explicit and computable expression, and $A(\lambda) = \sum_{\bfP} \abs{B(\bfP,\lambda)}$. Here, $\lambda$ and $C$ are variational parameters, in which $\lambda$ is a set of parameters that determine the distribution, and $C = \sum_{\rm \bfP} \abs{q(\bfP)}$ is a real number that represents the overhead cost of the error mitigation. 

Let $f$ be the measurement outcome of the quantum circuit specified by $\bfR$ and $\bfP$, and its distribution is ${\rm Pro}(f\vert\bfR,\bfP)$. Then, the computing result, i.e.~the mean value of $f$, reads 
\begin{eqnarray}
{\rm com}(\bfR,\bfP) = \sum_f {\rm Pro}(f\vert\bfR,\bfP) f.
\end{eqnarray}
Similarly, the error-free computing result can be expressed as 
\begin{eqnarray}
{\rm com}^{\rm ef}(\bfR,\bfP) = \sum_f {\rm Pro}^{\rm ef}(f\vert\bfR,\bfP) f.
\end{eqnarray}
In the Monte Carlo summation, the distribution ${\rm Pro}(f\vert\bfR,\bfP)$ is realised using the quantum computer, and all other distributions, including ${\rm Pro}^{\rm ef}(f\vert\bfR,\bfP)$, are realised on the classical computer. 

\subsection{Significant-error parametrisation}
\label{app:SigE}

We consider the first case. The error-mitigated computing result reads 
\begin{eqnarray}
{\rm com}^{\rm em}(\bfR,\bfI) = \sum_{\bfP\in{\rm SigE}} q(\bfP) {\rm com}(\bfR,\bfP).
\end{eqnarray}
Now, we consider the loss function, which is 
\begin{eqnarray}
{\rm Loss} &=& \frac{1}{\abs{\mathbb{T}}} \sum_{\bfR\in\mathbb{T}}\vert {\rm com}^{\rm em}(\bfR,\bfI)-{\rm com}^{\rm ef}(\bfR,\bfI)\vert^{2} \notag \\
&=& \sum_{\bfP,\bfP'\in{\rm SigE}} a_{\bfP,\bfP'} q(\bfP)q(\bfP') \notag \\
&&- 2\sum_{\bfP\in{\rm SigE}} b_{\bfP} q(\bfP) + c
\end{eqnarray}
where 
\begin{eqnarray}
a_{\bfP,\bfP'} &=& \frac{1}{\abs{\mathbb{T}}} \sum_{\bfR} {\rm com}(\bfR,\bfP){\rm com}(\bfR,\bfP'), \\
b_{\bfP} &=& \frac{1}{\abs{\mathbb{T}}} \sum_{\bfR} {\rm com}(\bfR,\bfP){\rm com}^{\rm ef}(\bfR,\bfI), \\
c &=& \frac{1}{\abs{\mathbb{T}}} \sum_{\bfR} {\rm com}^{\rm ef}(\bfR,\bfI)^2.
\end{eqnarray}

The optimal quasi-probability distribution is 
\begin{eqnarray}
\overline{q}_{\rm opt} = \overline{a}^{-1} \overline{b},
\end{eqnarray}
where $\overline{q}$ is a $\abs{\rm SigE}$-dimensional column vector with the elements $q(\bfP)$, $\overline{a}$ is a $\abs{\rm SigE}$-dimensional matrix with the elements $a_{\bfP,\bfP'}$, and $\overline{b}$ is a $\abs{\rm SigE}$-dimensional column vector with the elements $b_{\bfP}$. The minimum value of the loss function is 
\begin{eqnarray}
{\rm Loss}_{\rm min} = c - \overline{b}^{\rm T} \overline{a}^{-1} \overline{b}.
\end{eqnarray}

\subsubsection{The computation of $a_{\bfP,\bfP'}$}

We have 
\begin{eqnarray}
a_{\bfP,\bfP'} &=& \frac{1}{\abs{\mathbb{T}}} \sum_{\bfR,f,f'} {\rm Pro}(f\vert\bfR,\bfP) {\rm Pro}(f'\vert\bfR,\bfP') f f'.~~~~~
\end{eqnarray}
To compute $a_{\bfP,\bfP'}$, we generate independent and identically distributed samples $\{(\bfR_i,f_i,f_i')\vert i = 1,2,\ldots,N_{\rm s}\}$ according to the distribution 
\begin{eqnarray}
{\rm Pro}(\bfR) {\rm Pro}(f\vert\bfR,\bfP) {\rm Pro}(f'\vert\bfR,\bfP'), \notag
\end{eqnarray}
where 
\begin{eqnarray}
{\rm Pro}(\bfR) &=& \frac{1}{\abs{\mathbb{T}}}.
\end{eqnarray}
The estimator of $a_{\bfP,\bfP'}$ is 
\begin{eqnarray}
\hat{a}_{\bfP,\bfP'} = \frac{1}{N_{\rm s}} \sum_{i=1}^{N_{\rm s}} f_i f'_i.
\end{eqnarray}
The variance of the estimator is 
\begin{eqnarray}
{\rm Var}\left[\hat{a}_{\bfP,\bfP'}\right] = \frac{1}{N_{\rm s}}{\rm Var}\left[f f'\right].
\end{eqnarray}
Let $\abs{f}_{\rm max}$ be the maximum value of $\abs{f(\bfmu)}$, we have $f f' \leq \abs{f}_{\rm max}^2$. Therefore, 
\begin{eqnarray}
{\rm Var}\left[\hat{a}_{\bfP,\bfP'}\right] \leq \frac{1}{N_{\rm s}} \abs{f}_{\rm max}^4.
\end{eqnarray}

\subsubsection{The computation of $b_{\bfP}$}

We have 
\begin{eqnarray}
b_{\bfP} &=& \frac{1}{\abs{\mathbb{T}}} \sum_{\bfR,f,f'} {\rm Pro}(f\vert\bfR,\bfP) {\rm Pro}^{\rm ef}(f'\vert\bfR,\bfI) f f'.~~
\end{eqnarray}
To compute $b_{\bfP}$, we generate independent and identically distributed samples $\{(\bfR_i,f_i,f_i')\vert i = 1,2,\ldots,N_{\rm s}\}$ according to the distribution 
\begin{eqnarray}
{\rm Pro}(\bfR) {\rm Pro}(f\vert\bfR,\bfP) {\rm Pro}^{\rm ef}(f'\vert\bfR,\bfI). \notag
\end{eqnarray}
The estimator of $b_{\bfP}$ is 
\begin{eqnarray}
\hat{b}_{\bfP} = \frac{1}{N_{\rm s}} \sum_{i=1}^{N_{\rm s}} f_i f'_i.
\end{eqnarray}
The variance of the estimator is 
\begin{eqnarray}
{\rm Var}\left[\hat{b}_{\bfP}\right] = \frac{1}{N_{\rm s}}{\rm Var}\left[f f'\right] \leq \frac{1}{N_{\rm s}} \abs{f}_{\rm max}^4.
\end{eqnarray}

\subsubsection{The computation of $c$}

We have 
\begin{eqnarray}
c &=& \frac{1}{\abs{\mathbb{T}}} \sum_{\bfR,f,f'} {\rm Pro}^{\rm ef}(f\vert\bfR,\bfP) {\rm Pro}^{\rm ef}(f'\vert\bfR,\bfI) f f'.~~
\end{eqnarray}
To compute $c$, we generate independent and identically distributed samples $\{(\bfR_i,f_i,f_i')\vert i = 1,2,\ldots,N_{\rm s}\}$ according to the distribution 
\begin{eqnarray}
{\rm Pro}(\bfR) {\rm Pro}^{\rm ef}(f\vert\bfR,\bfP) {\rm Pro}^{\rm ef}(f'\vert\bfR,\bfI). \notag
\end{eqnarray}
The estimator of $c$ is 
\begin{eqnarray}
\hat{c} = \frac{1}{N_{\rm s}} \sum_{i=1}^{N_{\rm s}} f_i f'_i.
\end{eqnarray}
The variance of the estimator is 
\begin{eqnarray}
{\rm Var}\left[\hat{c}\right] = \frac{1}{N_{\rm s}}{\rm Var}\left[f f'\right] \leq \frac{1}{N_{\rm s}} \abs{f}_{\rm max}^4.
\end{eqnarray}

\subsubsection{The computation of ${\rm Loss}_{\rm min}$}

The estimator of ${\rm Loss}_{\rm min}$ is 
\begin{eqnarray}
\hat{\rm Loss}_{\rm min} = \hat{c} - \hat{\overline{b}}^{\rm T} \hat{\overline{a}}^{-1} \hat{\overline{b}}.
\end{eqnarray}
The variance of the estimator is 
\begin{eqnarray}
&& {\rm Var}\left[\hat{\rm Loss}_{\rm min}\right] \notag \\
&\simeq & {\rm E}\left[ \left( \delta c + \hat{\overline{b}}^{\rm T} \hat{\overline{a}}^{-1} \delta\overline{a} \hat{\overline{a}}^{-1} \hat{\overline{b}} - 2\hat{\overline{b}}^{\rm T} \hat{\overline{a}}^{-1} \delta\hat{\overline{b}} \right)^2 \right] \notag \\
&\simeq & {\rm Var}\left[\hat{c}\right] + \sum_{\bfP,\bfP'} q_{\rm opt}(\bfP)^2 {\rm Var}\left[\hat{a}_{\bfP,\bfP'}\right] q_{\rm opt}(\bfP')^2 \notag \\
&&+ 4\sum_{\bfP} q_{\rm opt}(\bfP)^2 {\rm Var}\left[\hat{b}_{\bfP}\right] \notag \\
&\leq & \frac{1}{N_{\rm s}} \abs{f}_{\rm max}^4 (1 + \abs{\overline{q}_{\rm opt}}^4 + 4\abs{\overline{q}_{\rm opt}}^2),
\end{eqnarray}
where $\delta\overline{a} = \hat{\overline{a}} - \overline{a}$, $\delta\overline{b} = \hat{\overline{b}} - \overline{b}$, $\delta c = \hat{c} - c$ and $\abs{\overline{q}_{\rm opt}}^2 = \sum_{\bfP} q_{\rm opt}(\bfP)^2$. The overhead cost of the error mitigation is $C = \sum_{\bfP} \abs{q_{\rm opt}(\bfP)}$. Because $C^2\geq \abs{\overline{q}_{\rm opt}}^2$, we have 
\begin{eqnarray}
{\rm Var}\left[\hat{\rm Loss}_{\rm min}\right] \lesssim \frac{1}{N_{\rm s}} \abs{f}_{\rm max}^4 (1 + C^4 + 4C^2).
\end{eqnarray}

\subsubsection{The computation of ${\rm com}^{\rm em}$}

Given the optimal quasi-probability distribution $q_{\rm opt}(\bfP)$, we can implement the error-mitigated computing accordingly. Taking $q(\bfP) = q_{\rm opt}(\bfP)$, we have 
\begin{eqnarray}
&& {\rm com}^{\rm em}(\bfR,\bfI) \notag \\
&=& \sum_{\bfP,f} q(\bfP) {\rm Pro}(f\vert\bfR,\bfP) f \notag \\
&=& \sum_{\bfP,f} \frac{\abs{q(\bfP)}}{C} {\rm Pro}(f\vert\bfR,\bfP) C\frac{q(\bfP)}{\abs{q(\bfP)}}f.
\end{eqnarray}
To compute ${\rm com}^{\rm em}(\bfR,\bfI)$, we generate independent and identically distributed samples $\{(\bfP_i,f_i)\vert i = 1,2,\ldots,N_{\rm s}\}$ according to the distribution 
\begin{eqnarray}
{\rm Pro}(\bfP) {\rm Pro}(f\vert\bfR,\bfP), \notag
\end{eqnarray}
where 
\begin{eqnarray}
{\rm Pro}(\bfP) = \frac{\abs{q(\bfP)}}{C}.
\end{eqnarray}
The estimator of ${\rm com}^{\rm em}(\bfR,\bfI)$ is 
\begin{eqnarray}
\hat{\rm com}^{\rm em}(\bfR,\bfI) = \frac{1}{N_{\rm s}} \sum_{i=1}^{N_{\rm s}} C\frac{q(\bfP_i)}{\abs{q(\bfP_i)}}f_i.
\end{eqnarray}
The variance of the estimator is 
\begin{eqnarray}
{\rm Var}\left[\hat{\rm com}^{\rm em}(\bfR,\bfI)\right] &=& \frac{1}{N_{\rm s}}{\rm Var}\left[C\frac{q(\bfP)}{\abs{q(\bfP)}}f\right] \notag \\
&\leq & \frac{1}{N_{\rm s}} \abs{f}_{\rm max}^2 C^2.
\end{eqnarray}

\subsection{General parametrisation}

We consider the second case. The error-mitigated computing result reads 
\begin{eqnarray}
{\rm com}^{\rm em}(\bfR,\bfI) = \sum_{\bfP} C \frac{B(\bfP,\lambda)}{A(\lambda)} {\rm com}(\bfR,\bfP).
\end{eqnarray}
Here, $\frac{\abs{B(\bfP,\lambda)}}{A(\lambda)}$ is a normalised distribution. Because the number of $\bfP$ grows exponentially with the number of single-qubit Pauli gates in the circuit, it could be difficult to compute the normalisation factor $A(\lambda)$ given the explicit expression of $B(\bfP,\lambda)$. Although we may not be able to compute $A(\lambda)$, we can sample the distribution $\frac{\abs{B(\bfP,\lambda)}}{A(\lambda)}$ using the Metropolis method. 

Now, we consider the loss function, which depends on $\lambda$ and $C$, i.e. 
\begin{eqnarray}
{\rm Loss}(C,\lambda) &=& \frac{1}{\abs{\mathbb{T}}} \sum_{\bfR\in\mathbb{T}}\vert {\rm com}^{\rm em}(\bfR,\bfI)-{\rm com}^{\rm ef}(\bfR,\bfI)\vert^{2} \notag \\
&=& aC^2 - 2bC + c,
\end{eqnarray}
where 
\begin{eqnarray}
a &=& \frac{1}{\abs{\mathbb{T}}} \sum_{\bfR,\bfP,\bfP'} \frac{B(\bfP,\lambda)B(\bfP',\lambda)}{A(\lambda)^2} \notag \\
&&\times {\rm com}(\bfR,\bfP){\rm com}(\bfR,\bfP'), \\
b &=& \frac{1}{\abs{\mathbb{T}}} \sum_{\bfR,\bfP} \frac{B(\bfP,\lambda)}{A(\lambda)} {\rm com}(\bfR,\bfP){\rm com}^{\rm ef}(\bfR,\bfI).
\end{eqnarray}

Our purpose is to minimise the loss function and find the optimal $C$ and $\lambda$. Given the quadratic form of the loss function, the optimal value of $C$ is 
\begin{eqnarray}
C_{\rm opt} = \frac{b}{a},
\end{eqnarray}
and the corresponding minimum value of the loss function is 
\begin{eqnarray}
{\rm Loss}_{\rm min}(\lambda) &=& c-\frac{b^2}{a},
\end{eqnarray}
which is still a function of $\lambda$. We note that $c-\frac{b^2}{a} \geq 0$ is always true. 

We can find that, in expressions of $a$ and $b$, coefficients are normalised distributions. Therefore, we can compute $a$ and $b$ using the Monte Carlo summation and generate samples using the Metropolis method. 

\subsubsection{Importance sampling}

Usually, only a small subset of Pauli errors are dominant. Accordingly, the optimal solution $q(\bfP)$ is only significant for a small subset of error-mitigating gate sequences, and $q(\bfP)$ is close to zero for most of $\bfP$. Therefore, if the variance of $f$ is finite, generating samples according to $q(\bfP)$ (i.e.~$B(\bfP,\lambda)$) is sub-optimal for the Monte Carlo summation. 

We evaluate the loss function in order to find the optimal distribution. Usually, we need to actively update the distribution $q(\bfP)$ (i.e.~$\lambda$ and $C$). For efficiently utilising the samples, we will need to use the samples generated according to the distribution $B(\bfP,\lambda')$, which is close to $B(\bfP,\lambda)$, to compute $a$ and $b$. Then, it is not necessary to generate new samples every time when we update $\lambda$. 

\subsubsection{The computation of $a$}

We have 
\begin{eqnarray}
a &=& \frac{1}{\abs{\mathbb{T}}} \sum_{\bfR,\bfP,\bfP',f,f'} \frac{B(\bfP,\lambda)B(\bfP',\lambda)}{A(\lambda)^2} \notag \\
&&\times {\rm Pro}(f\vert\bfR,\bfP) {\rm Pro}(f'\vert\bfR,\bfP') f f' \notag \\
&=& \frac{A(\lambda')^2}{A(\lambda)^2} \sum_{\bfR,\bfP,\bfP',f,f'} \frac{1}{\abs{\mathbb{T}}} \frac{\abs{B(\bfP,\lambda')B(\bfP',\lambda')}}{A(\lambda')^2} \notag \\
&&\times {\rm Pro}(f\vert\bfR,\bfP) {\rm Pro}(f'\vert\bfR,\bfP') \notag \\
&&\times \frac{B(\bfP,\lambda)B(\bfP',\lambda)}{\abs{B(\bfP,\lambda')B(\bfP',\lambda')}} f f' \notag \\
&=& \frac{A(\lambda')^2}{A(\lambda)^2} \tilde{a}.
\end{eqnarray}
To compute $\tilde{a}$, we generate independent and identically distributed samples $\{(\bfR_i,\bfP_i,\bfP_i',f_i,f_i')\vert i = 1,2,\ldots,N_{\rm s}\}$ according to the distribution 
\begin{eqnarray}
{\rm Pro}(\bfR) {\rm Pro}(\bfP) {\rm Pro}(\bfP') {\rm Pro}(f\vert\bfR,\bfP) {\rm Pro}(f'\vert\bfR,\bfP'), \notag
\end{eqnarray}
where 
\begin{eqnarray}
{\rm Pro}(\bfP) &=& \frac{\abs{B(\bfP,\lambda')}}{A(\lambda')}.
\end{eqnarray}
The estimator of $\tilde{a}$ is 
\begin{eqnarray}
\hat{\tilde{a}} = \frac{1}{N_{\rm s}} \sum_{i=1}^{N_{\rm s}} \frac{B(\bfP_i,\lambda)B(\bfP'_i,\lambda)}{\abs{B(\bfP_i,\lambda')B(\bfP'_i,\lambda')}} f_i f'_i.
\end{eqnarray}
The variance of the estimator is 
\begin{eqnarray}
{\rm Var}\left[\hat{\tilde{a}}\right] = \frac{1}{N_{\rm s}}{\rm Var}\left[\frac{B(\bfP,\lambda)B(\bfP',\lambda)}{\abs{B(\bfP,\lambda')B(\bfP',\lambda')}} f f'\right].
\end{eqnarray}
If $\lambda' = \lambda$, we have 
\begin{eqnarray}
{\rm Var}\left[\hat{\tilde{a}}\right] \leq \frac{1}{N_{\rm s}} \abs{f}_{\rm max}^4.
\end{eqnarray}

\subsubsection{The computation of $b$}

We have 
\begin{eqnarray}
b &=& \frac{1}{\abs{\mathbb{T}}} \sum_{\bfR,\bfP,f,f'} \frac{B(\bfP,\lambda)}{A(\lambda)} \notag \\
&&\times {\rm Pro}(f\vert\bfR,\bfP) {\rm Pro}^{\rm ef}(f'\vert\bfR,\bfI) f f' \notag \\
&=& \frac{A(\lambda')}{A(\lambda)} \sum_{\bfR,\bfP,f,f'} \frac{1}{\abs{\mathbb{T}}} \frac{\abs{B(\bfP,\lambda')}}{A(\lambda')} \notag \\
&&\times {\rm Pro}(f\vert\bfR,\bfP) {\rm Pro}^{\rm ef}(f'\vert\bfR,\bfI) \notag \\
&&\times \frac{B(\bfP,\lambda)}{\abs{B(\bfP,\lambda')}} f f' \notag \\
&=& \frac{A(\lambda')}{A(\lambda)} \tilde{b}.
\end{eqnarray}
To compute $\tilde{b}$, we generate independent and identically distributed samples $\{(\bfR_i,\bfP_i,\bfP_i',f_i,f_i')\vert i = 1,2,\ldots,N_{\rm s}\}$ according to the distribution 
\begin{eqnarray}
{\rm Pro}(\bfR) {\rm Pro}(\bfP) {\rm Pro}(f\vert\bfR,\bfP) {\rm Pro}^{\rm ef}(f'\vert\bfR,\bfI). \notag
\end{eqnarray}
The estimator of $\tilde{b}$ is 
\begin{eqnarray}
\hat{\tilde{b}} = \frac{1}{N_{\rm s}} \sum_{i=1}^{N_{\rm s}} \frac{B(\bfP_i,\lambda)}{\abs{B(\bfP_i,\lambda')}} f_i f'_i.
\end{eqnarray}
The variance of the estimator is 
\begin{eqnarray}
{\rm Var}\left[\hat{\tilde{b}}\right] = \frac{1}{N_{\rm s}}{\rm Var}\left[\frac{B(\bfP,\lambda)}{\abs{B(\bfP,\lambda')}} f f'\right].
\end{eqnarray}
If $\lambda' = \lambda$, we have 
\begin{eqnarray}
{\rm Var}\left[\hat{\tilde{b}}\right] \leq \frac{1}{N_{\rm s}} \abs{f}_{\rm max}^4.
\end{eqnarray}

\subsubsection{The computation of ${\rm Loss}_{\rm min}$}

We have 
\begin{eqnarray}
{\rm Loss}_{\rm min}(\lambda) = c - \frac{b^2}{a} = c - \frac{\tilde{b}^2}{\tilde{a}}.
\end{eqnarray}
Therefore, the estimator of ${\rm Loss}_{\rm min}(\lambda)$ is 
\begin{eqnarray}
\hat{\rm Loss}_{\rm min}(\lambda) = \hat{c} - \frac{\hat{\tilde{b}}^2}{\hat{\tilde{a}}}.
\end{eqnarray}
The variance of the estimator is 
\begin{eqnarray}
&& {\rm Var}\left[\hat{\rm Loss}_{\rm min}(\lambda)\right] \notag \\
&\simeq & {\rm E}\left[ \left( \delta c + \frac{\tilde{b}^2}{\tilde{a}^2}\delta\tilde{a} - \frac{2\tilde{b}}{\tilde{a}}\delta\tilde{b} \right)^2 \right] \notag \\
&\simeq & {\rm Var}\left[\tilde{c}\right] + \frac{\tilde{b}^4}{\tilde{a}^4}{\rm Var}\left[\tilde{\bar{a}}\right] + \frac{4\tilde{b}^2}{\tilde{a}^2}{\rm Var}\left[\hat{\tilde{b}}\right],
\end{eqnarray}
where $\delta\tilde{a} = \hat{\tilde{a}} - \tilde{a}$ and $\delta\tilde{b} = \hat{\tilde{b}} - \tilde{b}$. If $\lambda' = \lambda$, we have $C_{\rm opt} = \frac{\tilde{b}}{\tilde{a}}$ and 
\begin{eqnarray}
{\rm Var}\left[\hat{\rm Loss}_{\rm min}(\lambda)\right] \lesssim \frac{1}{N_{\rm s}} \abs{f}_{\rm max}^4(1+C_{\rm opt}^4+4C_{\rm opt}^2).~~~~~
\end{eqnarray}

\subsubsection{Computation of ${\rm com}^{\rm em}$}

By minimising ${\rm Loss}_{\rm min}(\lambda)$, we can obtain the optimal value of $\lambda$, which is $\lambda_{\rm opt}$. Then, the optimal quasi-probability distribution is given by $\lambda_{\rm opt}$ and the corresponding $C_{\rm opt}$, and we can implement the error-mitigated computing accordingly. We remark that, to compute $C_{\rm opt} = \frac{b}{a}$, we need to generate samples with $\lambda' = \lambda$, then $a = \tilde{a}$ and $b = \tilde{b}$. Taking $\lambda = \lambda_{\rm opt}$ and $C = C_{\rm opt}$, we have 
\begin{eqnarray}
{\rm com}^{\rm em}(\bfR,\bfI) &=& \sum_{\bfP,f} C\frac{B(\bfP,\lambda)}{A(\lambda)} {\rm Pro}(f\vert\bfR,\bfP) f \notag \\
&=& \sum_{\bfP,f} \frac{\abs{B(\bfP,\lambda)}}{A(\lambda)} {\rm Pro}(f\vert\bfR,\bfP) \notag \\
&&\times C \frac{B(\bfP,\lambda)}{\abs{B(\bfP,\lambda)}} f.~~~
\end{eqnarray}
To compute ${\rm com}^{\rm em}(\bfR,\bfI)$, we generate independent and identically distributed samples $\{(\bfP_i,f_i)\vert i = 1,2,\ldots,N_{\rm s}\}$ according to the distribution 
\begin{eqnarray}
{\rm Pro}(\bfP) {\rm Pro}(f\vert\bfR,\bfP), \notag
\end{eqnarray}
where 
\begin{eqnarray}
{\rm Pro}(\bfP) &=& \frac{\abs{B(\bfP,\lambda)}}{A(\lambda)}.
\end{eqnarray}
The estimator of ${\rm com}^{\rm em}(\bfR,\bfI)$ is 
\begin{eqnarray}
\hat{\rm com}^{\rm em}(\bfR,\bfI) = \frac{1}{N_{\rm s}} \sum_{i=1}^{N_{\rm s}} C \frac{B(\bfP_i,\lambda)}{\abs{B(\bfP_i,\lambda)}} f_i.
\end{eqnarray}
The variance of the estimator is 
\begin{eqnarray}
{\rm Var}\left[\hat{\rm com}^{\rm em}(\bfR,\bfI)\right] &=& \frac{1}{N_{\rm s}}{\rm Var}\left[C \frac{B(\bfP,\lambda)}{\abs{B(\bfP,\lambda)}} f\right] \notag \\
&\leq & \frac{1}{N_{\rm s}} \abs{f}_{\rm max}^2 C^2.
\end{eqnarray}

\end{document}